\documentclass[nofootinbib,reprint,superscriptaddress,prC]{revtex4-2}

\usepackage{multirow}
\usepackage{siunitx}
\usepackage{tikz}
\usetikzlibrary{patterns}
\usetikzlibrary{decorations.pathmorphing}

\usepackage{graphicx,times}
\usepackage{latexsym}
\usepackage{mathtools}

\usepackage{amsmath,amssymb,amsbsy,amsfonts}
\usepackage{array}
\usepackage{graphics}
\usepackage{xcolor}
\usepackage{cancel}
\usepackage{xspace}
\usepackage{nicefrac}
\usepackage[normalem]{ulem}
\usepackage{dsfont}
\usepackage[withpage,nohyperlinks,nolist]{acronym} 
\usepackage[unicode, pdfprintscaling=None]{hyperref}
\hypersetup{
    colorlinks,
    linkcolor={red!50!black},
    citecolor={blue!50!black},
    urlcolor={blue!80!black}
}

\usepackage[capitalise]{cleveref}

\usepackage{xcolor}
\usepackage{makecell}

\newcommand{\vect}[1]{\vec{\boldsymbol{#1}}}

\newcommand{\Gb}{\boldsymbol{\mathcal{G}}}
\newcommand{\Xb}{\mathbf{X}}
\newcommand{\Mb}{\mathbf{M}}
\newcommand{\Ob}{\mathbf{O}}

\newcommand{\Kb}{\mathbf{K}}

\newcommand{\Nb}{\mathbf{N}}

\newcommand{\Bb}{\mathbf{B}}

\newcommand{\Db}{\mathbf{D}}
\newcommand{\Qb}{\mathbf{Q}}

\newcommand{\eftnopi}{EFT($\cancel{\pi}$)\xspace}

\newcommand{\triton}{{}^3\mathrm{H}}
\newcommand{\He}{{}^3\mathrm{He}}

\DeclareMathSymbol{\shortminus}{\mathbin}{AMSa}{"39}

\usepackage{orcidlink}

\usepackage{relsize}

\usepackage{etoolbox}

\makeatletter
\newcommand*{\rom}[1]{\expandafter\@slowromancap\romannumeral #1@}

\newcolumntype{H}{>{\setbox0=\hbox\bgroup}c<{\egroup}@{}}

\begin{document}
\title{Coulomb Effects and Wigner-SU(4) Symmetry in He-3 Charge and Magnetic Properties}

\author{Xincheng Lin\,\orcidlink{0000-0001-9068-6787}}
\email{xlin28@ncsu.edu}
\affiliation{Department of Physics, North Carolina State University, Raleigh, North Carolina 27607, USA}

\begin{abstract}
This work studies the non-perturbative Coulomb corrections to the $\He$ binding energy, magnetic moment, and charge and magnetic radii in \ac{LO} \ac{eftnopi}. The splitting between $\He$ and $\triton$ binding energy is found to be $0.85(3)$~MeV. The Coulomb corrections to the $\He$ point charge radius and full magnetic radius are found to be $0.043(2)$~fm and $0.036(2)$~fm, respectively. These corrections are $\approx 4\%$ of the \ac{LO} predictions without Coulomb and should be taken into account at \ac{NNLO} or beyond in \ac{eftnopi} to achieve the desired EFT accuracy. The Coulomb correction to the $\He$ magnetic moment is found to be $-0.0041(1)\mu_N$, only $\approx 0.2\%$ of the \ac{LO} prediction without Coulomb. The impact of Wigner-SU(4) symmetry in the presence of the non-perturbative Coulomb interaction is also discussed and used to help explain the hierarchy of Coulomb effects in $\He$ observables. 
\end{abstract}

\maketitle
\newpage
\acresetall

\section{Introduction}
\Ac{eftnopi} is an effective field theory that describes nuclear systems with a typical momentum, $Q$, well below the pion mass, $m_\pi \approx 140$~MeV (see Ref.~\cite{Hammer:2019poc} for a recent review on nuclear EFTs). In \ac{eftnopi}, nucleons are the only dynamic degrees of freedom and interact through short-range forces organized systematically in powers of $Q/m_\pi$. In the two-nucleon sector, \ac{eftnopi} at \ac{LO} reproduces the deuteron bound state and the spin-singlet virtual state, and at higher orders it perturbatively reproduces the effective range expansion.~\cite{Chen:1999tn}. External currents, such as electromagnetic currents, can be included to study  deuteron form factors and neutron-proton radiative capture~\cite{Chen:1999vd,Chen:1999bg,Rupak:1999rk}. 
\ac{eftnopi} has also been used extensively to study three-nucleon systems, such as neutron-deutron scattering ($nd$) and $\triton$~\cite{Gabbiani:1999yv,Bedaque:2002yg,Griesshammer:2004pe,Vanasse:2013sda,Bedaque:1998mb,Bedaque:1999ve,Bedaque:1999vb}, and proton-deuteron scattering ($pd$) and $\He$~\cite{Rupak:2001ci,Ando:2010wq,Koenig:2011lmm,Vanasse:2014kxa,Konig:2016iny,Konig:2013cia,Konig:2014ufa,Konig:2015aka}. Including external electromagnetic currents allows one to study, for example, the $\He$ and $\triton$ electromagnetic properties~\cite{Vanasse:2015fph,Vanasse:2017kgh,Platter:2005sj,Kirscher:2017fqc,Kirscher:2010dgl} and $nd$ radiative capture~\cite{Lin:2022yaf,Lin:2024bor,Sadeghi:2006fc,Arani:2014qsa}. Many of the observables mentioned above have been studied beyond \ac{NLO} in \ac{eftnopi}, where the naive EFT truncation error is about $(Q/m_\pi)^{m+1} \sim 0.3^{m+1}$ at N$^m$LO using the deuteron binding momentum $\gamma_t$ to estimate $Q \sim \gamma_t\approx 45.7$~MeV . 

For systems with two or more protons, the Coulomb interaction also occurs. The size of the Coulomb effect can be estimated using the ratio between the Coulomb momentum scale $\kappa_c$ and the momentum scale of the observable under consideration. In particular, the relevant Coulomb momentum for two protons is $ \kappa_c = \alpha M_N \approx 8$~MeV, where $\alpha\approx 1/137$ is the fine-structure constant and $M_N\approx938$~MeV is the nucleon mass. 
For observables characterized by a momentum close to or below $\kappa_c $, such as  proton-proton~\cite{Kong:1998sx,Kong:1999sf,Ando:2007fh} or proton-deuteron scattering~\cite{Vanasse:2014kxa,Konig:2013cia,Konig:2014ufa,Konig:2016iny,Koenig:2011lmm} at a momentum below $\kappa_c$, the Coulomb interaction should be included non-perturbatively.
In contrast, both perturbative and non-perturbative treatments of the Coulomb effect are viable for $\He$ static properties as the Coulomb correction is expected to be of order $\alpha M_N/\sqrt{M_N B^{\mathrm{exp}}_{\He}} \sim 8\%$, where $B^{\mathrm{exp}}_{\He} = 7.71$~MeV is the $\He$ binding energy. This is roughly of the same size as a \ac{NNLO} term in the \ac{eftnopi} expansion and suggests that it is necessary to include the Coulomb effect to obtain a complete description of $\He$ static properties at \ac{NNLO} or beyond in \ac{eftnopi}. The Coulomb correction to the $\He$ binding energy has been studied in \ac{eftnopi} with the Coulomb interaction treated non-perturbatively~\cite{Ando:2010wq}, partially non-perturbatively~\cite{Vanasse:2014kxa}, and fully perturbatively~\cite{Konig:2014ufa,Konig:2015aka,Konig:2016iny,HaAndJared2026}. On the other hand, while the $\He$ electromagnetic properties have been studied at \ac{NNLO} in the \ac{eftnopi} expansion~\cite{Vanasse:2015fph, Vanasse:2017kgh}, the Coulomb correction to these observables has not been discussed.

This work studies the non-perturbative Coulomb effects on $\He$ static properties, including the $\He$ binding energy, magnetic moment, and charge, magnetic, and matter radii. The purpose is three-fold. First, this work completes the gap in the existing studies of $\He$ static properties, where the Coulomb effect is omitted even at \ac{NNLO} in \ac{eftnopi}. This work obtains a $\sim 4 \%$ Coulomb correction to the $\He$ radii, which is important for a precise description of $\He$ static properties at \ac{NNLO} and beyond in \ac{eftnopi}.
Second, this study is a precursor to an \ac{eftnopi} study of $pd$ radiative capture into $\He$, an important reaction in nuclear astrophysics. 
The final proton-deuteron scattering state in this process requires a non-peturbative treatment of the Coulomb interaction. 
Recent experimental results of the $pd$ capture cross section and its angular distribution are given by Refs.~\cite{Cavanna:2018mdc,Stockel:2024hde,2019EPJA...55..137T}. In particular, Ref.~\cite{Stockel:2024hde} demonstrates a tension between the experimental result and the potential-model calculation~\cite{Marcucci:2015yla,Stockel:2024hde}, which is somewhat difficult to evaluate as the potential-model result does not come with a systematic error estimate. An EFT calculation would provide a benchmark and help understand the origin of this tension.\footnote{The author would like to thank Daniel R. Phillips for bringing this into the author's attention.}
Finally, this work studies the impact of  Wigner-SU(4) symmetry in $\He$ with the non-perturbative Coulomb interaction. 
 Wigner-SU(4) symmetry refers to the combined spin-isospin symmetry where the nucleon state is treated as a four-dimenional supermultiplet~\cite{Wigner:1936dx}.\footnote{The Wigner-SU(4) symmetry can be perceived as an ($S$-wave) accidental symmetry that emergies in the limit of $N_c\to \infty$, where $N_c$ is the number of colors in \ac{QCD}. For details, see Ref.~\cite{Kaplan:1995yg,Kaplan:1996rk,Lee:2020esp}.} Refs.~\cite{Vanasse:2016umz,Vanasse:2017kgh} have shown that, in the absence of Coulomb interactions,  Wigner-SU(4) symmetry is a good approximate symmetry for three-nucleon bound states, with a $\sim10\%$ Wigner-SU(4) breaking effect. 
The inclusion of the Coulomb interaction further breaks  Wigner-SU(4) symmetry and is expected to affect its role in $\He$. 
This work analyzes the interplay between  Wigner-SU(4) symmetry and the Coulomb interaction in $\He$ and quantifies the Wigner-SU(4) breaking effect in the presence of the non-perturbative Coulomb interaction.
This will also be useful for understanding the impact of  Wigner-SU(4) symmetry on $pd$ capture in the future, as  Wigner-SU(4) symmetry is known to affect the power counting and error analysis in  neutron-deuteron capture~\cite{Lin:2022yaf,Lin:2024bor}.

This paper is structured as follows. Sec.~\ref{sec:3N_bound_states} introduces the theoretical framework for two- and three-nucleon systems at \ac{LO} in \ac{eftnopi}  with non-perturbative Coulomb interaction and derives the $\He$ vertex and wavefunction. 
Sec.~\ref{sec:He3_observables} shows the expressions for $\He$ static properties, including the binding energy, magnetic moment, and charge, magnetic, and matter radii. 
Sec.~\ref{sec:Wigner} analyzes  Wigner-SU(4) symmetry and its breaking in $\He$ in the presence of non-perturbative Coulomb interaction.
Sec.~\ref{sec:results} presents the numerical results for the $\He$ observables and discusses the interplay between  Wigner-SU(4) symmetry and the Coulomb interaction. 
Finally, a summary and outlook are given in Sec.~\ref{sec:conclusion}.

\section{Three-nucleon bound states in Pionless EFT}
\label{sec:3N_bound_states}
\subsection{Lagrangian and two-body sector}
The Lagrangian in \ac{eftnopi} up to three nucleons can be written as
\begin{align}
    \label{eq:Lagrangian_full}
    \mathcal{L}=&\hat{N}^{\dagger}\left(iD_0+\frac{\vect{\mathbf{D}}^2}{2M_N}\right)\hat{N}  + \mathcal{L}_{2} + \mathcal{L}_{3}+ \mathcal{L}_{\text{photon}},
\end{align}
where $\hat{N}$ is the nucleon field, $\mathcal{L}_{2}$ and  $\mathcal{L}_{3}$ contain two- and three-body interactions, respectively, and the $\mathcal{L}_{\text{photon}}$ contains the photon Lagrangian. The covariant derivative for nucleons is defined by
\begin{equation}
    D_\mu=\partial_\mu+ie\frac{1+\tau_3}{2}\hat{A}_\mu.
\end{equation}
The static Coulomb propagator is given by
\begin{align}
    iD_\gamma(\vect{q}) = \frac{i}{\vect{q}^2 + m_{\gamma}^2},
\end{align}
where $q$ is the photon three-momentum and $m_{\gamma}$ is a finite photon mass used to regulate infrared divergences. 
\begin{figure}[htb!]
    \centering
    \includegraphics[width=0.4\textwidth]{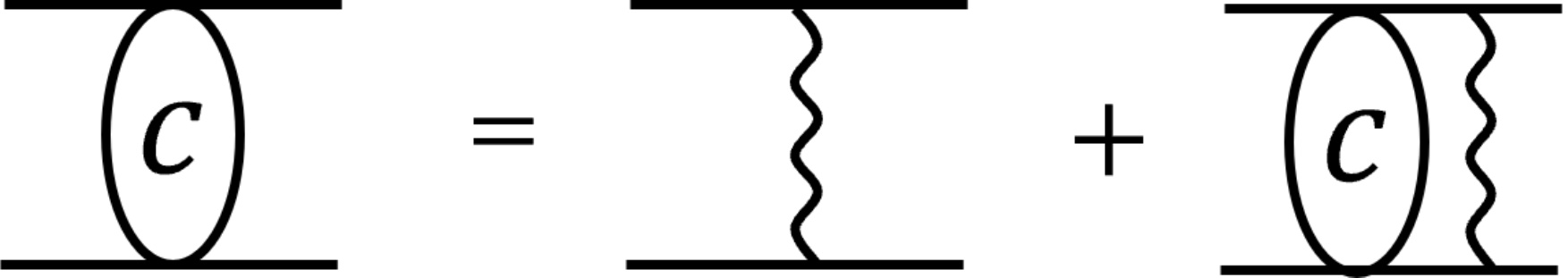}
    \caption{Coulomb T-matrix in the form of an integral equation. Solid and wavy lines represent nucleons and Coulomb photons, respectively.}
    \label{fig:Coulomb_T_matrix}
\end{figure}
The integral equation for the \ac{pp} Coulomb $T$-matrix $t_c(E;\vect{k},\vect{p})$ is shown diagrammatically in Fig.~\ref{fig:Coulomb_T_matrix}. It can be projected onto a partial-wave basis, 
\begin{align}
\label{eq:tc_partial_wave}
    t_c(E;\vect{k},\vect{p})  = \sum_{\ell=0}^{\infty} (2\ell + 1)t_c^{\ell}(E;k,p)P_\ell(\hat{\boldsymbol{k}}\cdot\hat{\boldsymbol{p}})
\end{align}
and
\begin{align}
    \label{eq:tc_ell}
    t_c^{\ell}(E;k,p) = &\frac{1}{2}\int_{-1}^{1}d(\hat{\boldsymbol{k}}\cdot\hat{\boldsymbol{p}})\,P_\ell(\hat{\boldsymbol{k}}\cdot\hat{\boldsymbol{p}}) t_c(E;\vect{k},\vect{p}),
\end{align}
where $P_\ell$ are Legendre polynomials. In practice the summation of $\ell$ in Eq.~\eqref{eq:tc_partial_wave} is truncated at an angular momentum cutoff $\ell_{\mathrm{max}}$.
$t_c^{\ell}(E;k,p)$ can be obtained numerically by  solving the integral equation in Fig.~\ref{fig:Coulomb_T_matrix}\footnote{An alternative method is to use the integral form of the Coulomb $T$-matrix in the limit $m_{\gamma}\to 0$ and subtract the infrared singularity analytically, as done in Ref.~\cite{Ando:2010wq}.}
\begin{align}
    t^\ell_c(E;{k},{p}) =& -4\pi\alpha D^\ell_\gamma(k,p)   \nonumber\\
    &+ \int_0^{\Lambda}\frac{q^2d{q}}{2\pi^2} 
    \frac{4\pi\alpha D^\ell_\gamma(k,q)}{E - \frac{q^2}{M_N} + i\varepsilon}t^\ell_c(E;{q},{p}),
\end{align}
where $\alpha$ is the fine-structure constant, $\Lambda$ is a sharp momentum cutoff, and $D^\ell_\gamma(k,p)$ is the $\ell$-th partial wave of the photon propagator
\begin{align}
   D^\ell_\gamma(k,p) =&\int \frac{d(\hat{\boldsymbol{k}}\cdot \hat{\boldsymbol{p}})}{2}D_\gamma(\vect{k} - \vect{p})P_\ell(\hat{\boldsymbol{k}}\cdot \hat{\boldsymbol{p}}) \nonumber\\
   =& -\frac{1}{2kp}Q_\ell\!\left(-\frac{k^2+p^2+m_{\gamma}^2}{2kp}\right),
\end{align}
with
\begin{align}
    Q_\ell\!\left(x\right) = \int_{-1}^1\frac{du}{2}\frac{P_\ell(u)}{u+x}.
\end{align}

\begin{figure*}[]
    \centering
    \includegraphics[width=0.8\textwidth]{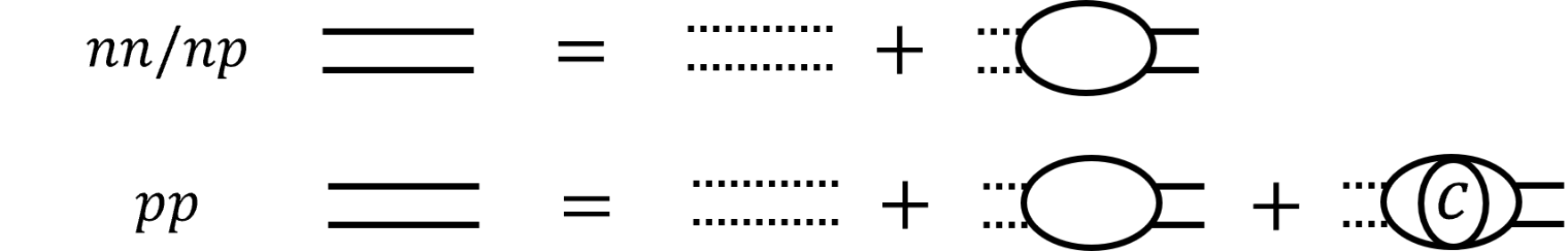}
    \caption{Dibaryon propagator in the form of an integral equation. The solid [dashed] double  line represents the dressed [bare] dibaryon propagator. }
    \label{fig:dibaryon_propagator}
\end{figure*}

Using the auxiliary field formalism, the \ac{LO} two-nucleon Lagrangian can be written as 
\begin{align}
\label{eq:Lagrangian_2body}
    \mathcal{L}_{2} =& \hat{t}_I^{\dagger}\Delta_t\hat{t}_I+\hat{s}_A^{\dagger}\Delta_s\hat{s}_A + \hat{s}_{+1}^{\dagger}\Delta_{pp}\hat{s}_{+1}\nonumber\\
 &+\sqrt{\frac{4\pi}{M_N}}\left(\hat{t}_I^{\dagger}\hat{N}^TP_I\hat{N} + \hat{s}_A^{\dagger}\hat{N}^T\bar{P}_A\hat{N} +\mathrm{H.c.}\right),
\end{align}  
where $\hat{t}_I$ [$\hat{s}_A$] is the spin-triplet [spin-singlet] dibaryon field and  $P_I=\frac{1}{\sqrt{8}}\sigma_2\sigma_I\tau_2$ [$\bar{P}_A=\frac{1}{\sqrt{8}}\sigma_2\tau_2\tau_A$] projects out the spin-triplet iso-singlet [spin-singlet] channel. $\hat{s}_{+1}$ is the dibaryon field for the \ac{pp} channel and $\Delta_{pp}$ is required to renormalize the Coulomb interaction. 
The dibaryon propagators can be obtained by solving analytically the integral equations in Fig.~\ref{fig:dibaryon_propagator}. 
The two-body \acp{LEC} $\Delta_t$ and $\Delta_s$ can be fit to reproduce the two-body binding momentum
\begin{align}
    \Delta_{t/s} &= \gamma_{t/s} - \frac{2\Lambda}{\pi},
\end{align}
where $\gamma_t = 45.7025$ MeV and $\gamma_s = -7.890$ MeV are the binding momentum of the deuteron and the spin-singlet virtual state, respectively. (An alternative regularization is the power divegence subtraction scheme~\cite{Kaplan:1998tg} instead of a finite cutoff.)
The \ac{LO} dibaryon propagator in the limit $\Lambda\to\infty$ is given by~\cite{Bedaque:1999vb}
\begin{align}
    \label{eq:dibaryon_prop}
    D_{t/s}(E,p) &= \frac{1}{\gamma_{t/s} - \sqrt{\frac{p^2}{4} - M_NE - i\varepsilon}}, 
\end{align}
For the \ac{pp} channel, the two-body \ac{LEC} $\Delta_{pp}$ can be determined by~\cite{Konig:2015aka}
\begin{align}
        \Delta_{pp} + \Delta_{s} &= \frac{1}{a_C} - \frac{2\Lambda}{\pi} + \alpha M_N\left(\ln\frac{2\Lambda}{\alpha M_N} - C_E\right),
\end{align}
where
$a_C = -7.8063$ fm is the Coulomb-modified \ac{pp} scattering length and $C_E \approx 0.5772$ is the Euler-Mascheroni constant. The \ac{LO} \ac{pp} propagator in the limit $\Lambda\to\infty$ is given by~\cite{Kong:1999sf,Ando:2010wq,Konig:2015aka}
\begin{align}
    D_{pp}(E,p) &= \frac{1}{\frac{1}{a_C} + \alpha M_NH(\eta)},  
\end{align}
where
\begin{align}
\label{eq:eta}
    \eta &= -i\frac{\alpha M_N}{2}\left(\frac{p^2}{4} - M_NE - i\varepsilon\right)^{-1/2},
\end{align}
and
\begin{align}
    H(\eta) &= \psi(i\eta) + \frac{1}{2i\eta} - \ln(i\eta),
\end{align}
with $\psi$ being the digamma function. 

Finally, a three-nucleon force is required at \ac{LO} to maintain  \ac{RG} invariance~\cite{Bedaque:1999ve}. The three-body Lagrangian can be written as
\begin{align}
\label{eq:L_3body}
    \mathcal{L}_{3} =& \hat{\psi}^{\dagger}\Omega\hat{\psi}+\sqrt{\frac{4\pi}{M_N}}\left(\hat{\psi}^{\dagger}\sigma_I\hat{N}\hat{t}_I-\hat{\psi}^\dagger\tau_A\hat{N}\hat{s}_A+\mathrm{H.c.}\right),
\end{align}
where $\hat{\psi}$ is the three-nucleon auxiliary field. The three-body \ac{LEC} $\Omega$ can be determined by fitting to the triton binding energy (see Sec.~\ref{sec:He3_observables} for how this is done).

\subsection{Helium-3 vertex function}
\begin{figure*}
    \centering
    \includegraphics[width=0.9\textwidth]{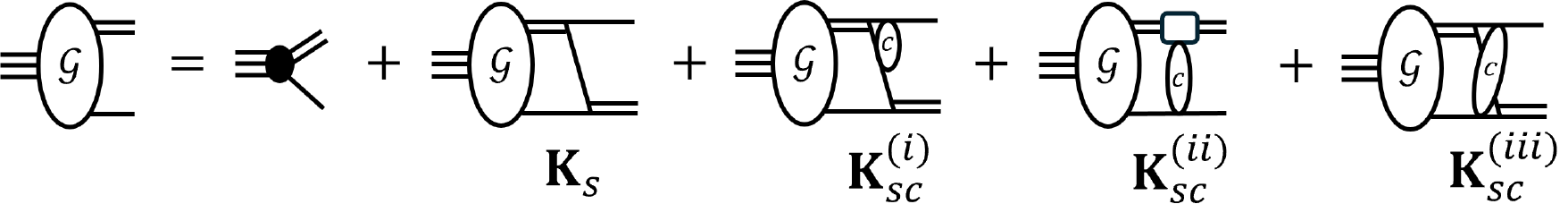}
    \caption{Integral equation for the $\He$ vertex function including non-perturbative Coulomb interaction~\cite{Ando:2010wq}. Triple line represents the three nucleon auxiliary field. Solid black circle represents the bare three-nucleon vertex. The bubble labeled $\mathcal{G}$ represents the $\He$ vertex function. Other notations are the same as Fig.~\ref{fig:Coulomb_T_matrix} and Fig.~\ref{fig:dibaryon_propagator}.}
    \label{fig:Helium3Vertex} 
\end{figure*}

Fig.~\ref{fig:Helium3Vertex} shows the integral equation for the $\He$ vertex function including the non-perturbative Coulomb interaction. 
The equation can be written compactly as 
\begin{align}
\label{eq:vertex_function}
    \Gb(E,p) &=  \Bb + \int_0^\Lambda \frac{q^2dq}{2\pi^2}\Kb(E,p,q) \Gb(E,q),
\end{align}
where $\Gb(E,p)$ and $\Bb$ are three-component vectors in  \ac{CCS}~\cite{Griesshammer:2004pe},
\begin{align}
    \label{eq:3B_vertex_cc_space}
    \Gb(E,p) &=
    \begin{pmatrix}
        \mathcal{G}_t(E,p)\\
        \mathcal{G}_s(E,p)\\
        \mathcal{G}_{pp}(E,p)
    \end{pmatrix}, \qquad
    \Bb =
    \begin{pmatrix}
        \sqrt{3}\\
       -1\\
       -\sqrt{2}
    \end{pmatrix},
\end{align}
where the subscripts $t$, $s$, and \ac{pp} denote the spin-triplet, spin-singlet (excluding \ac{pp}), and \ac{pp} channels, respectively. The inhomogeneous term $\Bb$ is obtained by writing the three-body force in the three-channel formalism using the normalized spin-doublet projector (see App.~\ref{app:cc_matrices} for details)
\begin{align}
\label{eq:3Ndoublet-projector}
\mathbf{P} = 
    \begin{pmatrix}
    {\sigma_I}/{\sqrt{3}} &  & \\
    & \tau_A\delta_{A3} & \\
    &&\tau_A\delta_{A\pm}
    \end{pmatrix},
\end{align}
where $\delta_{A\pm}$ is defined as$\sum_A\tau_A\delta_{A+} = (\tau_{1} \pm i\tau_{2})/\sqrt{2}$. 
The kernel, $\Kb(E,p,q)$, contains four contributions~\cite{Ando:2010wq},
\begin{align}
    \Kb(E,p,q) = &\Kb_{s}(E,p,q) + \Kb^{(\text{i})}_{sc}(E,p,q)  \nonumber\\
    &+ \Kb^{(\text{ii})}_{sc}(E,p,q) + \Kb^{(\text{iii})}_{sc}(E,p,q),
\end{align}
where $\Kb_{s}$ is the piece with pure strong interactions. $\Kb^{(\text{i})}_{sc}$, $\Kb^{(\text{ii})}_{sc}$, and $\Kb^{(\text{iii})}_{sc}$ involve both the strong and  Coulomb interaction, corresponding to the labeled diagrams in Fig.~\ref{fig:Helium3Vertex}. The explicit expression of $\Kb_{s}(E,p,q)$ in  three-channel formalism is
\begin{align}
    \label{eq:kernel_strong}
    \Kb_{s}(E,p,q) = & -\frac{16\pi}{pq}Q_0\!\left(\frac{p^2 + q^2 - M_NE}{pq}\right)\nonumber\\
    &\times
     \Xb \Db_d\!\left(E - \frac{q^2}{2M_N}, q\right) ,
\end{align}
where $\Db_d(E,q)$ is the three-channel representation of the dibaryon propagator,
\begin{align}
    \Db_d(E,q) =& 
    \begin{pmatrix}
        D_t(E,q)& &\\
        &D_s(E,q)&\\
        &&D_{pp}(E,q)
    \end{pmatrix},
\end{align}
and $\Xb$ is a three-channel \ac{CCS}  matrix that encodes the spin-isospin structure,
\begin{align}
    \label{eq:Xb}
    \Xb =&  -\frac{1}{8}
    \begin{pmatrix}
        -1& \sqrt{3}&\sqrt{6}\\
        \sqrt{3}& 1&-\sqrt{2}\\
        \sqrt{6}&-\sqrt{2}& 0
    \end{pmatrix}.
\end{align}
Note that $\Xb$ in Eq.~\eqref{eq:Xb} is constructed to be symmetric using the normalized spin-doublet projector in Eq.~\eqref{eq:3Ndoublet-projector}. This projector also fixes the relative normalization of the three  components of $\Bb$ in Eq.~\eqref{eq:3B_vertex_cc_space}. While this choice of the relative normalization differs from those in, e.g., Refs.~\cite{Ando:2010wq,Vanasse:2014kxa,Konig:2014ufa,Konig:2015aka}, it renders the three-channel \ac{CCS} matrices of various operators symmetric and is thus convenient to work with (see App.~\ref{app:cc_matrices} for an explicit connection between symmetric and non-symmetric choices of $\Xb$).

While some integrals in the rest of the kernel could be performed analytically as done in Ref.~\cite{Ando:2010wq}, the resulting expression are still quite complicated. Considering numerical interpolations are needed later to evaluate even more challenging integrals for the $\He$ magnetic moments and radii, this work chooses to compute some integrals below numerically.
To begin with, $\Kb^{(\text{i})}_{sc}(E,p,q)$ can be written as
\begin{align}
    \label{eq:kernel_alpha}
    \Kb^{(\text{i})}_{sc}(E,p,q) &= 
     -\frac{16\pi}{pq} Q_0\!\left(\frac{p^2 + q^2 - M_NE}{pq}\right)\nonumber\\  \times\Xb  \Nb_c\!&\left(E - \frac{3q^2}{4M_N}, p\right)\Db_d\!\left(E - \frac{q^2}{2M_N}, q\right),
\end{align}
where $\Nb_c(E,p)$ is the two-nucleon Coulomb vertex function
\begin{align}
\label{eq:Nc}
    \Nb_c(E,p) = & -\boldsymbol{\delta}_{33}\int_0^\Lambda \frac{l^2dl}{2\pi^2} t^{0}_c(E;{p},{l})\frac{1}{E - \frac{l^2}{M_N}}, 
\end{align}
with $\boldsymbol{\delta}_{33} =\text{diag}(0,0,1)$. $t^{0}_c$ is the $S$-wave Coulomb $T$-matrix defined in Eq.~\eqref{eq:tc_ell}. Note that Eq.~\eqref{eq:kernel_alpha} is almost the same as Eq.~\eqref{eq:kernel_strong}, except for the insertion of the two-nucleon Coulomb vertex function, $\Nb_c(E,p)$.

The third piece of the kernel, $\Kb^{(\text{ii})}_{sc}(E,p,q)$, can be written as
\begin{align}
    \label{eq:kernel_beta}
    \Kb^{(\text{ii})}_{sc}(E,p,q) = & 
    -L(E,p,q) \nonumber\\ 
    &\times\Xb_\beta \Db_d\!\left(E - \frac{q^2}{2M_N}. q\right),
\end{align}
where 
\begin{align}
\label{eq:Xb_beta}
\Xb_\beta = \begin{pmatrix}
        {1}/{4}& &\\
        & {1}/{4}&\\
        &&0
    \end{pmatrix},
\end{align}
with zero elements neglected. $L(E,p,q)$ is the subloop including the Coulomb $T$-matrix. 
\begin{figure}[]
    \centering
    \includegraphics[width=0.45\textwidth]{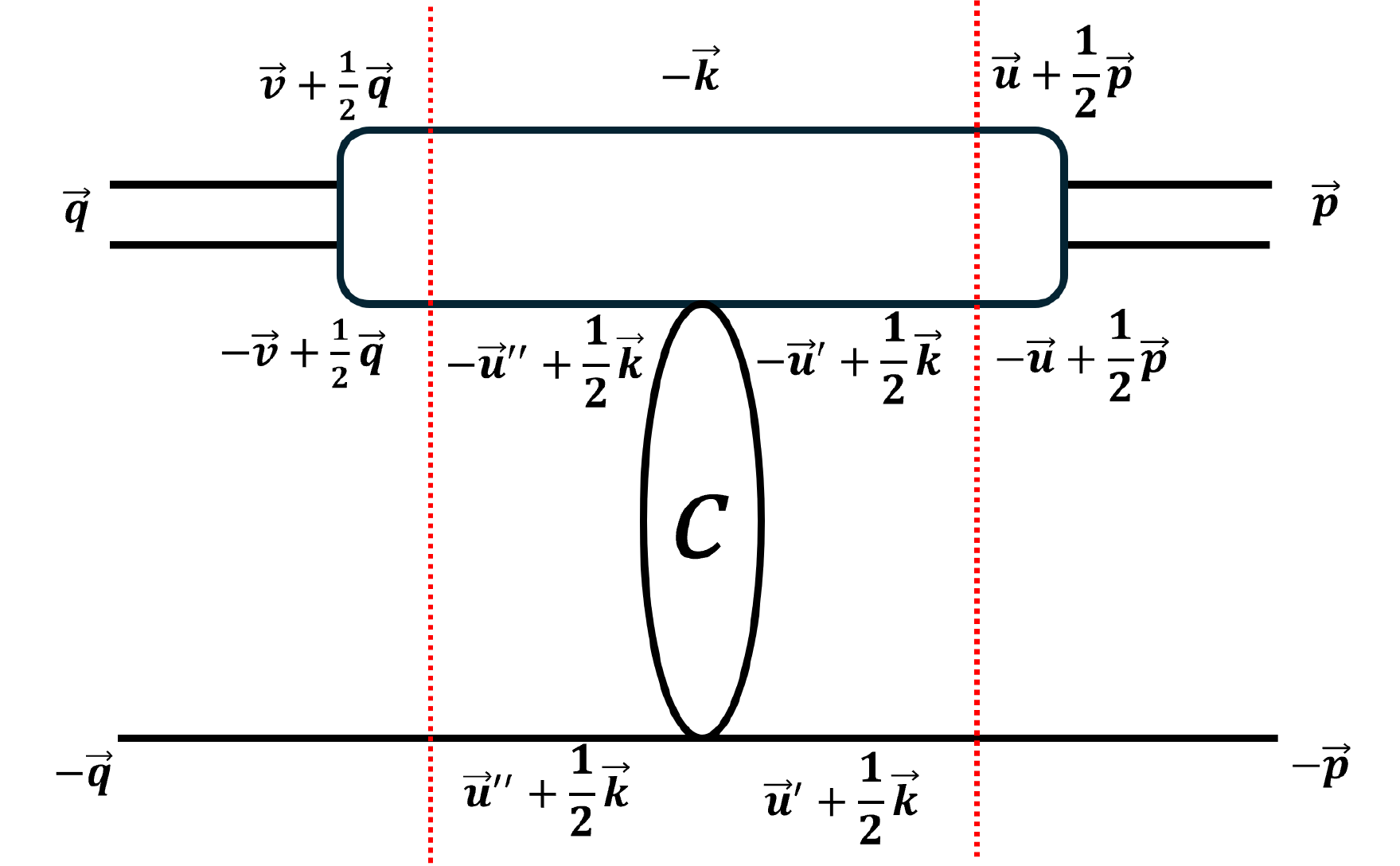}
    \caption{Diagram for $L(E,p,q)$ with partitions (vertical dashed red lines) of Jacobi momenta.}
    \label{fig:3B_Block}
\end{figure}
The diagram for $L(E,p,q)$ is shown in Fig.~\ref{fig:3B_Block} along with Jacobi momenta assigned for numerical evaluation. The expression for $L(E,p,q)$ can be obtained using the Coulomb $T$-matrix in a partial-wave basis,
\begin{align}
    L(E,p,q) = \sum_{\ell=0}^{\ell_{\mathrm{max}}} L_\ell(E,p,q),
\end{align}
where 
\begin{align}
    \label{eq:L_ell_interp}
    L_\ell(E,p,q) = & -\frac{16\pi}{M_N}\int \frac{u^2du}{2\pi^2}\frac{v^2dv}{2\pi^2}\frac{k^2dk}{2\pi^2}\frac{u'^2du'}{2\pi^2}\frac{u''^2du''}{2\pi^2} \nonumber\\
    &\times \frac{t^\ell_c\!\left(E - \frac{3k^2}{4M_N} ;{u}',{u}''\right)}{\left(E - \frac{3p^2}{4M_N} - \frac{u^2}{M_N}\right)\left(E - \frac{3k^2}{4M_N}  - \frac{u''^2}{M_N}  \right)  }  \nonumber\\
    &\times S^T_{0,\ell}( {u}, {p};{u}', {k})S^T_{\ell,0}({u}'', {k}; {v}, {q})\frac{2\pi^2}{q^2}.
\end{align}
On the last line above, $S^T_{\ell\ell'}(q_1, q_2;q_1',q_2')$ is the partial-wave projected interpolating function that connects two sets of Jacobi momenta in adjacent partitions in Fig.~\ref{fig:3B_Block},\footnote{$S^T_{\ell\ell'}$ in Eq.~\eqref{eq:interpolating-pw-main} is the (interpolated) overlap of the states defined in Eq.~\eqref{eq:pw_state} whose Jacobi momenta are related by the transformation matrix $T$ in Eq.~\eqref{eq:T-Jacobi}.} 
\begin{align}
\label{eq:interpolating-pw-main}
    S^T_{\ell\ell'}(q_1, q_2;q_1',q_2') &= \frac{1}{2}\int^{1}_{-1}d(\hat{\boldsymbol{q}}_1\cdot\hat{\boldsymbol{q}}_2)\tilde{P}_\ell\left(\hat{\boldsymbol{q}}_1\cdot\hat{\boldsymbol{q}}_2\right) \nonumber\\
    &\times \tilde{P}_{\ell'}\left(\widehat{(T\vect{Q})_1}\cdot\widehat{(T\vect{Q})_2}\right)\nonumber\\ 
    &\times S\left(|(T\vect{Q})_1|, |(T\vect{Q})_2|;q_1',q_2'\right),
\end{align}
where $\vect{Q} = (\vect{q}_1, \vect{q}_2)^\intercal$ and
\begin{align}
    \tilde{P}_\ell(\vect{q}_1, \vect{q}_2) = (-1)^\ell\sqrt{2\ell+1}{P}_\ell(\vect{q}_1, \vect{q}_2).
\end{align}
$T$ is the transformation matrix that connects two sets of the three-body Jacobi momentum in adjacent partitions in Fig.~\ref{fig:3B_Block},
\begin{align}
\label{eq:T-Jacobi}
    T = T^{-1} = \begin{pmatrix*}[r]
        {1}/{2}&{3}/{4}\\
        1 & -{1}/{2}
    \end{pmatrix*}.
\end{align}
$S(q_1,q_2;q'_1,q'_2)$ in Eq.~\eqref{eq:interpolating-pw-main} only involves the magnitude of the momenta and smears out products of Dirac delta functions
\begin{align}
    S(q_1,q_2;q'_1,q'_2) = 
   &C(q_1,q'_1)C(q_2,q'_2),\nonumber\\
    \sim &\delta(q'_1-q_1)\delta(q'_2-q_2),
\end{align}
In this work, $C(q_1,q'_1)$ is chosen to be a cubic spline. More details regarding interpolation can be found in App.~\ref{app:interpolation}.

The last piece of the kernel, $\Kb^{(\text{iii})}_{sc}(E,p,q)$, has exactly the same structure as $\Kb^{(\text{ii})}_{sc}(E,p,q)$ up to a permutation of the two nucleons connected to the Coulomb $T$-matrix. It can be written as
\begin{align}
    \label{eq:kernel_gamma}
    \Kb^{(\text{iii})}_{sc}(E,p,q) &=  
    -\sum_{\ell=0}^{\ell_{\mathrm{max}}} (-1)^\ell L_\ell(E, p, q)\nonumber\\ 
    &\times \Xb_\gamma \Db_d\!\left(E - \frac{q^2}{2M_N}. q\right),
\end{align}
where
\begin{align}
\label{eq:Xb_gamma}
    \Xb_\gamma = &  -\frac{1}{8}
    \begin{pmatrix}
        -1& \sqrt{3}&0\\
        \sqrt{3}& 1&0\\
        0&0& 0
    \end{pmatrix}.
\end{align}

\subsection{Helium-3 wavefunction}
In principle, $\He$ observables can be calculated using a diagrammatic approach with the $\He$ vertex function. However, with the Coulomb interaction this can be challenging and tedious due to the number and complexity of the form-factor diagrams involving Coulomb. An equivalent (and perhaps more tractable) approach is to first construct the $\He$ wavefunction from the vertex function, and then use the wavefunction to compute observables. The relation between the three-body vertex and  wavefunctions in the presence of the Coulomb interaction is derived in App.~\ref{app:vertex2wave} for three charged particles. The application to three-nucleon systems is straightforward by including the two-body spin-isospin projectors defined in the Lagrangian in Eq.~\eqref{eq:Lagrangian_2body}.

The full $\He$ wavefunction, denoted $|\boldsymbol{\Psi}\rangle$, can be written as a sum of two components
\begin{align}
\label{eq:Psi}
    |\boldsymbol{\Psi}\rangle = Z\frac{1}{3}(1 + \hat{\mathcal{P}})\left(|\boldsymbol{\psi}_{sc}\rangle + |\boldsymbol{\psi}_c\rangle\right),
\end{align}
where $Z$ is a normalization factor (see discussion below Eq.~\eqref{eq:psi-O-psi-0} for the computation of $Z$) and
\begin{align}
    \hat{\mathcal{P}} = \hat{\mathcal{P}}_{12}\hat{\mathcal{P}}_{23} + \hat{\mathcal{P}}_{31}\hat{\mathcal{P}}_{23},
\end{align}
with $\hat{\mathcal{P}}_{ij}$ interchanging the $i$-th and $j$-th particles and $\hat{\mathcal{P}}$ generating the three cyclic permutations for three particles.

\begin{figure*}[htb!]
    \centering
    \includegraphics[width=0.6\linewidth]{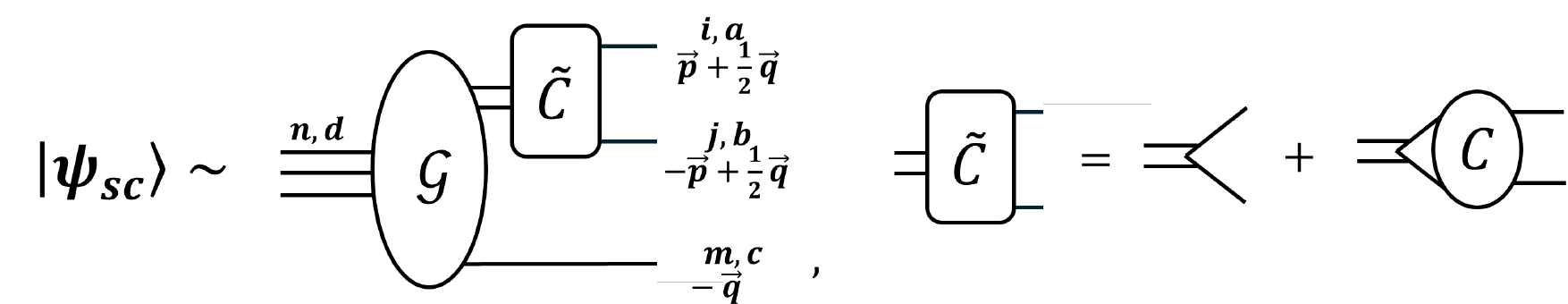}
    \caption{Schematic visualization of $|\boldsymbol{\psi}_{sc}\rangle$ with momentum, spin, and isospin indices correponding to those in Eqs.~\eqref{eq:psi_sc_he3} and~\eqref{eq:psi_sc_he3_tilde}. The  $\widetilde{C}$ bubble corresponds to the $1 + \Nb_c$ term in Eq.~\eqref{eq:psi_sc_he3_tilde}.}
    \label{fig:psi_sc}
\end{figure*}
\textit{First component of $|\boldsymbol{\Psi}\rangle$}. The first component in Eq.~\eqref{eq:Psi}, $|\boldsymbol{\psi}_{sc}\rangle$, is visualized in Fig.~\ref{fig:psi_sc} and has the representation 
\begin{align}
\label{eq:psi_sc_he3}
    \left(\boldsymbol{\psi}_{sc}\right)^{abc,d}_{ijm,n}(\vect{p},\vect{q}) =& \langle \vect{p},\vect{q};\{\sigma\}|\boldsymbol{\psi}_{sc}\rangle
    \nonumber\\ =&\Ob^{abc,d}_{ijm,n}\widetilde{\boldsymbol{\psi}}_{sc}(\vect{p},\vect{q})
\end{align}
where the indices ``$ijm$'' [``$abc$''] label the three-nucleon spins [isospins] and ``$n$'' [``$d$''] is the spin [isospin] index in the three-nucleon spin-$1/2$ [isospin-$1/2$] channel. 
In the bra-ket notation all the spin-isospin indices are contained in $\{\sigma\}$ for simplicity. $\Ob^{abc,d}_{ijm,n}$ selects the spin-$1/2$ and isospin-$1/2$ channel where $\He$ and $\triton$ live. In the three-channel formalism, $\Ob$ is given by
\begin{align}
    \Ob^{abc,d}_{ijm,n} = \begin{pmatrix}
        (\Ob_{11})^{abc,d}_{ijm,n} &  & \\
         & (\Ob_{22})^{abc,d}_{ijm,n} & \\
         &  & (\Ob_{33})^{abc,d}_{ijm,n}
    \end{pmatrix},
\end{align}
where the diagonal elements read
\begin{align}
    (\Ob_{11})^{abc,d}_{ijm,n} &=
        \left(P^\dagger_I\right)_{ij}^{ab}\frac{\left(\sigma_I\right)_{mn}}{\sqrt{3}}\delta_{cd} \\
    (\Ob_{22})^{abc,d}_{ijm,n}&=
        \left(\bar{P}^\dagger_3\right)_{ij}^{ab}\delta_{mn}(\tau_3)_{cd} \\
     (\Ob_{33})^{abc,d}_{ijm,n}&= \sum_{A}
        \left(\bar{P}^\dagger_A\right)_{ij}^{ab}\delta_{mn}(\tau_A)_{cd} \delta_{A+},
\end{align}
with $\delta_{A+} = (\delta_{A1}+ i\delta_{A2})/\sqrt{2}$. For $\He$ [$\triton$], the isospin index $d$ takes the upper [lower] component of the associated Pauli matrix or Kronecker delta in the isospin space, corresponding to an isospin of $+1/2$ [$-1/2$]. 
$\widetilde{\boldsymbol{\psi}}_{sc}(\vect{p},\vect{q})$ in Eq.~\eqref{eq:psi_sc_he3} is given by (see App.~\ref{app:vertex2wave} for a derivation)
\begin{align}
    \label{eq:psi_sc_he3_tilde}
    \widetilde{\boldsymbol{\psi}}_{sc}(\vect{p},\vect{q}) =& 
    \sqrt{\frac{4\pi}{M_N}}\frac{\left(1 + \Nb_c(E-\frac{3q^2}{4M_N},p)\right)}{-B_{\He} - \frac{3p^2}{4M_N} - \frac{q^2}{M_N}}\nonumber\\ &\times\Db_d\left(E-\frac{q^2}{2M_N},q\right)\Gb(-B_{\He} , q),
\end{align}
where $\Gb$ is the $\He$ vertex function in
Eq.~\eqref{eq:vertex_function} and $\Nb_c$ is defined in Eq.~\eqref{eq:Nc}. 

To facilitate the numerical evaluations of the $\He$ form factors in the later sections, it is useful to write Eqs.~\eqref{eq:psi_sc_he3} and~\eqref{eq:psi_sc_he3_tilde} in a partial-wave basis to remove the explicit angular dependence. 
The three-body state with the total angular momentum $L = 0$ has a complete partial-wave basis
\begin{align}
    \label{eq:pw_basis}
\mathbf{1}_{L=0} = \sum_{\ell,\ell'=0}^{\infty}\int_0^{\Lambda}\frac{p^2dp}{2\pi^2}\frac{q^2dq}{2\pi^2} |pq,(\ell\ell')\rangle\langle pq,(\ell\ell')|,
\end{align}
where 
\begin{align}
    \label{eq:pw_state}
    |pq,(\ell\ell')\rangle\equiv&|L = 0, M = 0,pq,(\ell\ell')\rangle \nonumber\\
    =&\sum_{m,m'} C_{\ell m, \ell' m'}^{00}|p,\ell m;q,\ell' m'\rangle \nonumber \\
    =&\frac{\delta_{\ell\ell'}}{4\pi}(-1)^{\ell} \sqrt{2\ell + 1}\int d\Omega_p d\Omega_q P_\ell(\hat{\boldsymbol{p}}\cdot\hat{\boldsymbol{q}})|\vect{p},\vect{q}\rangle.
\end{align}
In practice the summation over $\ell,\ell'$ in Eq.~\eqref{eq:pw_basis} is truncated at an angular momentum cutoff $\ell_{\mathrm{max}}$.
Finally,  $|\boldsymbol{\psi}_{sc}\rangle$ in this partial-wave basis has the expression
\begin{align}
    \langle pq,(\ell\ell');\{\sigma\}|\boldsymbol{\psi}_{sc}\rangle =& 4\pi\delta_{\ell0}\delta_{\ell'0}\Ob^{abc,d}_{ijm,n}\widetilde{\boldsymbol{\psi}}_{sc}(p,q),
\end{align}
where the angular dependence of $\widetilde{\boldsymbol{\psi}}_{sc}(\vect{p},\vect{q})$ is dropped as it only depends on the magnitudes of $\vect{p}$ and $\vect{q}$.

\begin{figure}[h]
    \centering
    \includegraphics[width=0.8\linewidth]{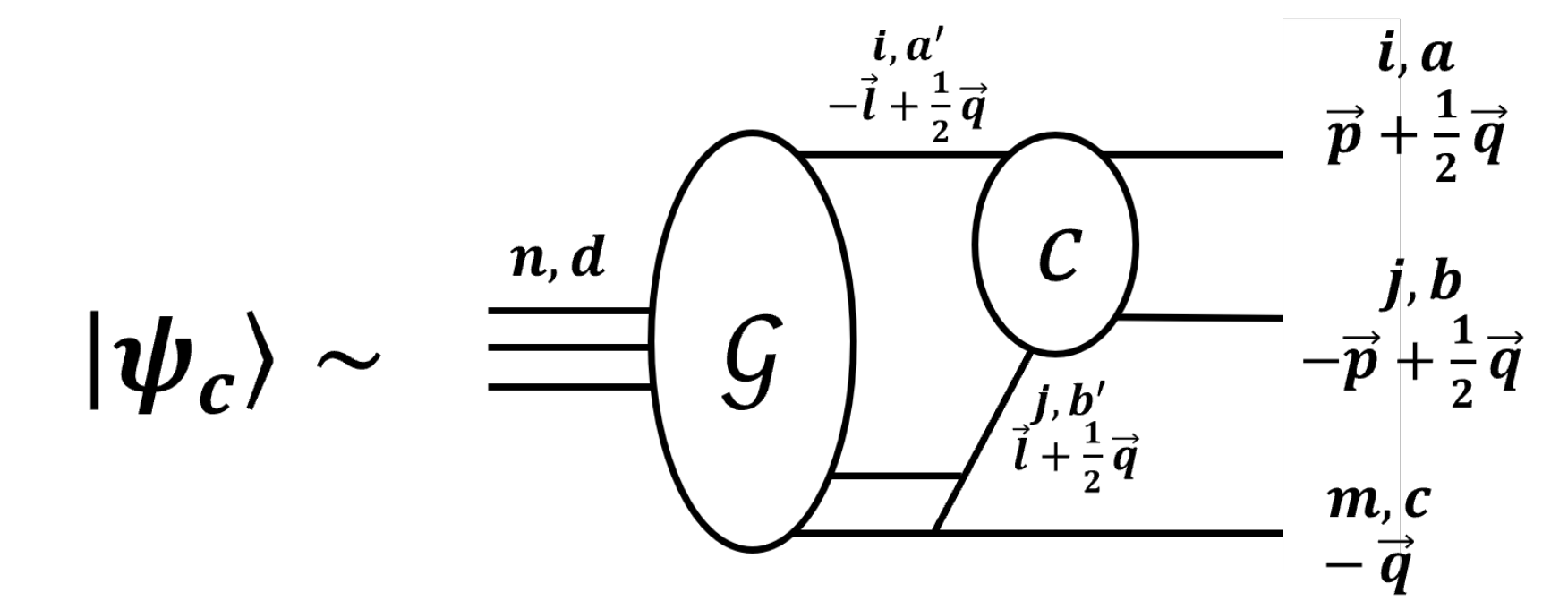}
    \caption{Schematic visualization of $|\boldsymbol{\psi}_{c}\rangle$ with momentum, spin, and isospin indices correponding to those in Eqs.~\eqref{eq:psi_c_he3} and~\eqref{eq:psi_c_he3_tilde}.}
    \label{fig:psi_c}
\end{figure}
\textit{Second component of $|\boldsymbol{\Psi}\rangle$}. The second component in Eq.~\eqref{eq:Psi}, $|\boldsymbol{\psi}_{c}\rangle$, is visualized in Fig.~\ref{fig:psi_c} and has the representation 
\begin{align}
\label{eq:psi_c_he3}
\left(\boldsymbol{\psi}_{c}\right)^{abc,d}_{ijm,n}(\vect{p}, \vect{q}) = &\langle \vect{p},\vect{q};\{\sigma\}|\boldsymbol{\psi}_{c}\rangle \nonumber\\
=&
\frac{1}{2}{(1+\tau_3)_{aa'}}\frac{1}{2}{(1+\tau_3)_{bb'}}\Ob^{cb'a',d}_{mji,n}\nonumber\\ &\qquad\qquad\qquad\qquad\times\widetilde{\boldsymbol{\psi}}_{c}(\vect{p},\vect{q}),
\end{align}
where repeated indices are summed over. $\widetilde{\boldsymbol{\psi}}_{c}(\vect{p},\vect{q})$ is related to $\widetilde{\boldsymbol{\psi}}_{sc}(\vect{p},\vect{q})$ through (see App.~\ref{app:vertex2wave} for a derivation)
\begin{align}
\label{eq:psi_c_he3_tilde}
    \widetilde{\boldsymbol{\psi}}_{c}(\vect{p}, \vect{q}) = &\int\frac{d^3\vect{l}}{(2\pi)^3}\frac{t_c(E;\vect{p}, \vect{l})}{-B_{\He} - \frac{3p^2}{4M_N} - \frac{q^2}{M_N}}\nonumber\\
    &\times \widetilde{\boldsymbol{\psi}}_{sc}\!\left(\frac{1}{2}\vect{l} + \frac{3}{4}\vect{q},\vect{l} - \frac{1}{2}\vect{q}\right).
\end{align}
where $\widetilde{\boldsymbol{\psi}}_{sc}\!\left(\frac{1}{2}\vect{l} + \frac{3}{4}\vect{q},\vect{l} - \frac{1}{2}\vect{q}\right)$ is the momentum-space component of $\langle\vect{l},\vect{q}| \hat{\mathcal{P}}_{12}\hat{\mathcal{P}}_{23}|\boldsymbol{\psi}_{sc}\rangle$. Using the partial-wave basis defined in Eq.~\eqref{eq:pw_state}, it can be expanded as
\begin{align}
\label{eq:psi_sc_boosted}
\widetilde{\boldsymbol{\psi}}_{sc}\!\left(\frac{1}{2}\vect{l} + \frac{3}{4}\vect{q},\vect{l} - \frac{1}{2}\vect{q}\right) = \sum_{\ell=0}^{\infty}&(-1)^{\ell}\sqrt{2\ell + 1}P_\ell(\hat{\boldsymbol{l}}\cdot\hat{\boldsymbol{q}})\nonumber\\ &\times\widetilde{\boldsymbol{\psi}}_{T,sc}^{\ell}(l,q),
\end{align}
where $\widetilde{\boldsymbol{\psi}}_{T,sc}^{\ell}(l,q)$ is the momentum space component of $\langle pq,(\ell\ell')| \hat{\mathcal{P}}_{12}\hat{\mathcal{P}}_{23}|\boldsymbol{\psi}_{sc}\rangle/4\pi$ and can be computed using the interpolation function in Eq.~\eqref{eq:interpolating-pw-main}, 
\begin{align}
    \widetilde{\boldsymbol{\psi}}_{T,sc}^{\ell}(l,q) = \int_0^\Lambda\frac{l'^2dl'}{2\pi^2}\frac{q'^2dq'}{2\pi^2}&S^T_{\ell0}(l, q;l',q')\widetilde{\boldsymbol{\psi}}_{sc}(l',q').
\end{align}
Using Eqs.~\eqref{eq:psi_c_he3_tilde},~\eqref{eq:psi_sc_boosted}, and~\eqref{eq:tc_partial_wave}, it is straightforward to write $|\psi_{c}\rangle$ in the partial-wave basis,
\begin{align}
\langle  pq,(\ell\ell');\{\sigma\}|\boldsymbol{\psi}_{c}\rangle 
=& \frac{1}{2}{(1+\tau_3)_{aa'}}\frac{1}{2}{(1+\tau_3)_{bb'}}\Ob^{cb'a',d}_{mji,n}\nonumber\\ &\times4\pi \delta_{\ell\ell'}\widetilde{\boldsymbol{\psi}}^\ell_{c}(p,q),
\end{align}
where   $\widetilde{\boldsymbol{\psi}}^\ell_{c}(p,q)$ is given by
\begin{align}
    \widetilde{\boldsymbol{\psi}}^\ell_{c}({p}, {q}) = \int_0^{\Lambda}\frac{l^2dl}{2\pi^2}&\frac{t_c^\ell(E;{p}, {l})}{-B_{\He} - \frac{3p^2}{4M_N} - \frac{q^2}{M_N}}\widetilde{\boldsymbol{\psi}}_{T,sc}^{\ell}(l,q).
\end{align}

\section{Helium-3 observables}
\label{sec:He3_observables}

\subsection{Helium-3 binding energy}
\begin{figure}[htb!]
    \centering
    \includegraphics[width=0.2\textwidth]{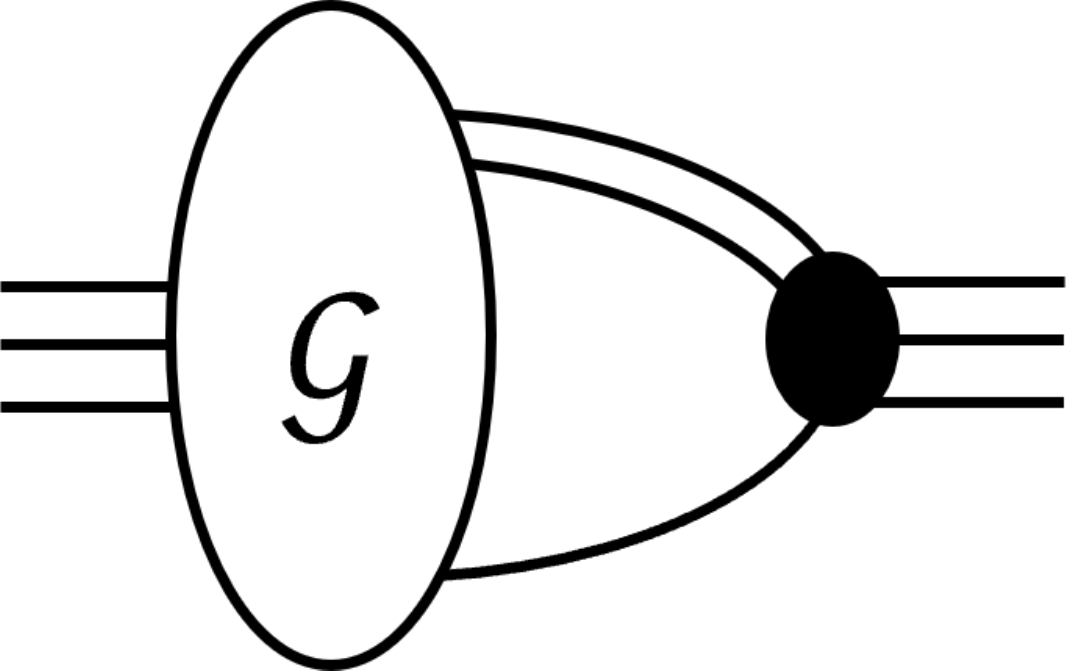}
    \caption{Trimer irreducible diagram.}
    \label{fig:onePI3body}
\end{figure}

One way to find  the $\He$ binding energy, $B_{\He}$, is to solve for the energy where the dressed $\He$ propagator hits a singularity. This is equivalent to solving (see Ref.~\cite{Vanasse:2015fph} for the triton case)
\begin{align}
\label{eq:B3He_condition}
    1 = H_0\Sigma_{\He}(-B_{\He}),
\end{align}
where $\Sigma_{\He}(E)$ is the trimer irreducible diagram in Fig.~\ref{fig:onePI3body} for $\He$,
\begin{align}
    \Sigma_{\He}(E) = &\int_0^\Lambda \frac{q^2dq}{2\pi^2}\Bb^\intercal\Db_d\!\left(E - \frac{q^2}{2M_N}, q\right)\Gb(E,q),
\end{align}
and
\begin{align}
    H_0 = -\frac{4\pi}{M\Omega}.
\end{align}
In this work, $H_0$ is fit at each cutoff $\Lambda$ to the $\triton$ binding energy, $B^{\textrm{exp}}_{\triton} = 8.48$~MeV, yielding the expression
\begin{align}
    H_0 = \frac{1}{\Sigma_{\triton}(-B^{\textrm{exp}}_{\triton})},
\end{align}
where $\Sigma_{\triton}(E)$ is the trimer irreducible diagram for $\triton$ and is defined in the same way as $\Sigma_{\He}(E)$ but with the Coulomb interaction turned off. Once $H_0$ is known from fitting, $B_{\He}$ can be obtained by solving Eq.~\eqref{eq:B3He_condition}.

\subsection{Helium-3 form factors}
\label{sec:He3form}
This section shows the calculation of the $\He$ form factors and radii using the $\He$ wavefunction. 
An operator $\hat{\rho}$ given by a sum of single-nucleon operators $\hat{\rho}_i(\vect{k})$ with a momentum transfer $\vect{k}$ can be written as
\begin{align}
    \hat{\rho} (\vect{k}) &= \sum_{i=1}^{A}\hat{\rho}_i(\vect{k}).
\end{align}
For a two-body system, $\hat{\rho}_i(\vect{k})$ in Jacobi momenta can be written as
\begin{align}
    \hat{\rho}_i(\vect{k}) = \int \frac{d^3\vect{p}}{(2\pi)^3}|\vect{p}+\tfrac{(-1)^i}{2}\vect{k}\rangle\langle \vect{p}|,
\end{align}
and for a three-body system,
\begin{align}
\label{eq:rho_i_3body}
    \hat{\rho}_i(\vect{k}) = \begin{cases}
        \int \frac{d^3\vect{p}}{(2\pi)^3}\frac{d^3\vect{q}}{(2\pi)^3}|\vect{p}+\tfrac{(-1)^i}{2}\vect{k},\vect{q}+\tfrac{\vect{k}}{3}\rangle\langle \vect{p},\vect{q}|.
        & i = 1,2\\
        \int \frac{d^3\vect{p}}{(2\pi)^3}\frac{d^3\vect{q}}{(2\pi)^3}|\vect{p},\vect{q}-\tfrac{2}{3}\vect{k}\rangle\langle \vect{p},\vect{q}|.
        & i = 3
    \end{cases}
\end{align}
The two-body matrix element of $\hat{\rho}_i(\vect{k})$ in the partial-wave basis is given by (see Refs.~\cite{Konig:2019xxk,Andis:2025fsg} for a detailed derivation.)
\begin{align}
\label{eq:two-body-rho-matrix-element}
\langle p, \ell m|&\hat{\rho}_i(\vect{k})|p', \ell' m'\rangle = \delta_{\ell \ell'}\delta_{mm'}\sum_{n} \sqrt{\binom{2\ell}{2n}} C^{\ell0}_{n0,(\ell-n)0} \nonumber\\
&\times \int_{-1}^{1} \frac{du}{2}\, P_n(u) \frac{\delta\left(p'- v(p, k, u)\right)}{p'^2/2\pi^2} \frac{p^{\ell-n}\left(-\frac{1}{2}k\right)^n}{v(p, k, u)^\ell},
\end{align}
where 
\begin{align}
   v(p, k, u)  = \sqrt{p^2 - pku + k^2/4}.
\end{align}
For an operator with a separable spin-isospin structure, it can be written as  
\begin{align}
    \hat{\rho}^\xi(\vect{k}) =& \sum_{i=1}^{A}\hat{\rho}^\xi_i(\vect{k}) \nonumber \\
    =& \sum_{i=1}^{A}\hat{\rho}_i(\vect{k})\otimes \hat{O}_i^\xi
\end{align}
where $\xi = \#, C, s, v, M$ labels the type of the operator in the spin-isospin space. $\hat{O}_i^\xi$ are given by
\begin{align}
    \hat{O}_i^\# = \mathbf{1}, &\qquad \hat{O}_i^C =\frac{1}{2}(\mathbf{1}+\tau_3)_i,
\end{align}
corresponding to the scalar and charge operator, respectively, and
\begin{align}
     \hat{O}^s = \kappa_0(\sigma_3)_i,\qquad \hat{O}^v=\kappa_1(\sigma_3\tau_3)_i, 
\end{align}
corresponding to the isoscalar and isovector magnetic operator, respectively.
$\kappa_0 = 0.4399 \mu_N$ and $\kappa_1 = 2.3529\mu_N$ are the isoscalar and isovector nucleon magnetic moments in nuclear magnetons ($\mu_N$), respectively. Finally,
\begin{align}
    {O}^M = \hat{O}^s + \hat{O}^v
\end{align}
corresponds to the full magnetic operator.
For simplicity, the magnetic operators above are written with only a $z$-component, and the $\He$ wavefunction below is assumed to be in the $|S=1/2, S_z = 1/2\rangle$ state.

\begin{figure*}[t]
    \centering
    \includegraphics[width=0.8\linewidth]{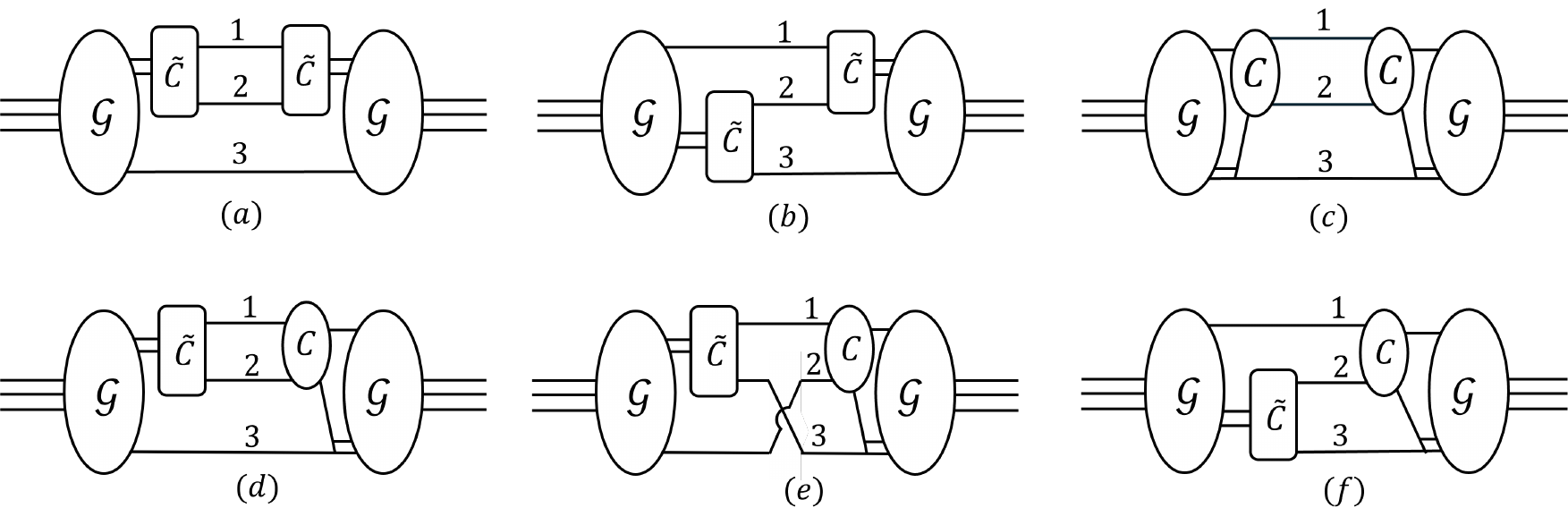}
    \caption{Schematic visualization of $\langle\Psi|\hat{\rho}|\Psi\rangle$ with all possible contractions of the numbered nucleons. The operator $\hat{\rho}_i$ acts on the $i$-th nucleon in the figure. These diagrams help track the spin-isospin summations (but are not used as Feynman diagrams in momentum space in this work).}
    \label{fig:He3formfactor}
\end{figure*}

The $\He$ form factor with $\hat{\rho}^\xi(\vect{k})$ can be written as
\begin{align}
\label{eq:psi-O-psi}
F^\xi_{\He}(k^2) =& \langle\Psi|\hat{\rho}^\xi(\vect{k})|\Psi\rangle\nonumber\\
= & Z\langle\boldsymbol{\psi}_{sc}|\hat{\rho}^\xi(\vect{k})\frac{1 + \hat{P}}{3}|\boldsymbol{\psi}_{sc}\rangle \nonumber\\
    &+  Z\langle\boldsymbol{\psi}_{c}|\hat{\rho}^\xi(\vect{k})\frac{1 + \hat{P}}{3}|\boldsymbol{\psi}_{c}\rangle\nonumber\\
    &+ Z\left(\langle\boldsymbol{\psi}_{c}|\hat{\rho}^\xi(\vect{k})\frac{1 + \hat{P}}{3}|\boldsymbol{\psi}_{sc}\rangle + \text{H.c.}\right),
\end{align}
where the fact that $1 + \hat{P}$ commutes with $\hat{\rho}^\xi(\vect{k})$ is used.
These matrix elements are visualized schematically in Fig.~\ref{fig:He3formfactor} (note that they are not used as Feynman diagrams in this work).
The form factor can be computed in momentum space by inserting Eq.~\eqref{eq:pw_basis} on each side of $\hat{\rho}^\xi(\vect{k})$ in Eq.~\eqref{eq:psi-O-psi}, yielding
\begin{align}
\label{eq:psi-O-psi-0}
    F^\xi_{\He}&(k^2) = Z\frac{1}{3}\int_0^\Lambda\frac{p^2dp}{2\pi^2}\frac{p'^2dp'}{2\pi^2}\frac{q^2dq}{2\pi^2}\frac{q'^2dq'}{2\pi^2}\nonumber\\
    \times& \sum_{\ell,\ell',\lambda, \lambda' = 0}^{\ell_{\mathrm{max}}} \sum_{i=1}^{3}\langle pq,(\ell\lambda)|\hat{\rho}_i(\vect{k})|p'q',(\ell'\lambda')\rangle \Bigg\{\nonumber\\
    &\widetilde{\boldsymbol{\psi}}_{sc}(p,q)\left[\Mb^\xi_{a,i}\widetilde{\boldsymbol{\psi}}_{sc}(p',q')+\Mb^\xi_{b,i}\widetilde{\boldsymbol{\psi}}^0_{T,sc}(p',q')\right]\delta_{\ell0}\nonumber\\
    +&\widetilde{\boldsymbol{\psi}}^\ell_{c}(p,q)\Mb^\xi_{c,i}\widetilde{\boldsymbol{\psi}}^\ell_{c}(p',q')\nonumber\\
    +&2\widetilde{\boldsymbol{\psi}}^\ell_{c}(p,q)\bigg[\Mb^\xi_{d,i}\widetilde{\boldsymbol{\psi}}_{sc}(p',q')\delta_{\ell0}+\Mb^\xi_{f,i}\widetilde{\boldsymbol{\psi}}^\ell_{T,sc}(p',q')\nonumber\\
    &\quad\qquad\qquad+(-1)^{\ell}\Mb^\xi_{e,i}\widetilde{\boldsymbol{\psi}}^\ell_{T,sc}(p',q')\bigg]\Bigg\},
\end{align}
where the three-body matrix element of $\hat{\rho}_i^\xi(\vect{k})$ is given by
\begin{align}
   \langle pq,(\ell\lambda)|\hat{\rho}_i(\vect{k})|p'q',(\ell'\lambda')\rangle
    &=\sum_{m,m', \nu,\nu'} C_{\ell m, \lambda \nu }^{00}C_{\ell' m', \lambda' \nu'}^{00} \nonumber \\
    &\hspace{-50pt}\times \langle p, \ell m; q,\lambda \nu |\hat{\rho}_i(\vect{k})|p', \ell' m'; q',\lambda' \nu'\rangle \nonumber \\
    &=\delta_{\ell\lambda}\delta_{\ell'\lambda'}\delta_{\lambda\lambda'}\rho_i^{\ell}(p,q,p',q',\vect{k}).
\end{align}
Using Eqs.~\eqref{eq:rho_i_3body} and~\eqref{eq:two-body-rho-matrix-element}, $\rho_i^{\ell}(p,q,p',q',\vect{k})$ can be written as
\begin{equation}
\label{eq:O_123}
    \rho_i^{\ell}(p,q,p',q',\vect{k}) =
    \begin{cases}
    \begin{aligned}
    &\langle p, \ell 0|\hat{\rho}_i(\vect{k})|p', \ell 0\rangle\\
    & \times \! \langle q, \ell 0|\hat{\rho}_1(\tfrac{2}{3}\vect{k})|q', \ell 0\rangle, \quad i = 1,2 
    \end{aligned} \\ \\
    \begin{aligned}
    &\frac{2\pi^2}{p^2}\delta(p'-p) \\
    &\times \!\langle q, \ell 0|\hat{\rho}_2(-\tfrac{4}{3}\vect{k})|q', \ell 0\rangle, \quad i = 3.
    \end{aligned}
    \end{cases}
\end{equation}
The two-matrix elements above are given by Eq.~\eqref{eq:two-body-rho-matrix-element}.
The fact that these matrix elements are diagonal in $\ell$ and $m$ has been used to simplify the expressions above.
The spin-isospin structure in Eq.~\eqref{eq:psi-O-psi-0} is encoded in
$\Mb^\xi_{\alpha,i}$ with $\alpha = a,b,\cdots$, which is the three-channel representation of the operator $\hat{\rho}_i$ with type-($\alpha$) contraction of the spin-isospin indices, as shown diagrammatically in Fig.~\ref{fig:He3formfactor}. The explicit expressions of  $\Mb^\xi_{\alpha,i}$ are given in App.~\ref{app:cc_matrices} for operators discussed above.
These expressions are calculated based on the contractions in Fig.~\ref{fig:He3formfactor} along with the spin-isospin component of the wavefunctions in Eqs.~\eqref{eq:psi_sc_he3} and~\eqref{eq:psi_c_he3}.  In particular, the normalization factor $Z$ in Eq.~\eqref{eq:Psi} is fixed by the condition $F^\#(0) = 3$. The isoscalar and isovector $\He$ magnetic moment are given by the form factor at zero momentum transfer,
\begin{align}
\label{eq:mu-from-FF}
\mu^{s/v}_{\He} = F^{s/v}_{\He}(0),
\end{align}
and the total $\He$ magnetic moment, $\mu_{\He}$ is given by their sum 
\begin{align}
    \mu_{\He} = \mu^s_{\He} + \mu^v_{\He}.
\end{align}

The $\He$ point radii are defined as
\begin{align}
\label{eq:radius-from-FF}
    \langle r_{0} \rangle^\xi_{\He} &\equiv \sqrt{\langle r^2_{0} \rangle^\xi_{\He}} \nonumber\\ 
    &= \sqrt{-\frac{6}{F^\xi_{\He}(0)}\frac{d}{d(k^2)}F^\xi_{\He}(k^2)\bigg|_{k=0}},
\end{align}
where the derivative with respect to $k^2$ can be computed using that of the two-body matrix element in Eq.~\eqref{eq:two-body-rho-matrix-element},
\begin{align}
\label{eq:two-body-rho-derivative}
\frac{d}{d(k^2)}\langle p, &\ell m|\hat{\rho}(\vect{k})|p', \ell' m'\rangle\bigg|_{k=0} = \delta_{\ell \ell'}\delta_{mm'}\left(\frac{p}{p'}\right)^{\ell}\frac{2\pi^2}{p'^2}\nonumber\\ & \times \frac{1}{6}\left[\frac{d^2}{d^2p}\delta(p'-p) + \frac{2(1+\ell)}{p}\frac{d}{dp}\delta(p'-p)\right].
\end{align}
Numerical treatment of the derivatives of delta functions is discussed in App.~\ref{app:interpolation}. Note that while $\rho_i^{\ell}$ in Eq.~\eqref{eq:O_123} for $i=1,2,3$ have the same form at $k=0$, their derivatives with respect to $k^2$ at $k=0$, obtained using Eq.~\eqref{eq:two-body-rho-derivative}, are different for $i=1,2$ and $i=3$. 

\section{Impact of the Wigner-SU(4) symmetry}
\label{sec:Wigner}
Wigner-SU(4) symmetry at \ac{LO} in \ac{eftnopi} can be parametrized by
\begin{align}
\gamma =& \frac{1}{2}\left(\gamma_t + \gamma_s\right), \qquad \delta = \frac{1}{2}\left(\gamma_t - \gamma_s\right),
\end{align}
where  $\gamma$ and $\delta$,  instead of $\gamma_t$ and $\gamma_s$, can also be used to parameterize
the dibaryon propagators in Eq.~\eqref{eq:dibaryon_prop}. Wigner-SU(4) symmetry is broken by a non-zero $\delta$ and the Wigner-SU(4) limit is defined by the limit $\delta \to 0$.
Previous studies~\cite{Vanasse:2016umz,Vanasse:2017kgh} have shown that, in the absence of Coulomb, three-nucleon bound states in the physical world are perturbatively close to the Wigner-SU(4) limit  with an expansion parameter ${\delta}/{\kappa^*_3}$, where $\kappa^*_3$ is a three-body momentum scale. Empirically these authors found this expansion parameter to be $\sim10\%$. Ref.~\cite{Vanasse:2017kgh} also studied the impact of  Wigner-SU(4) symmetry on the charge and magnetic properties of $\triton$ and $\He$. In particular, the \ac{LO} $\He$ magnetic moment in the Wigner-SU(4) limit is equal to the neutron magnetic moment (known as the Schmidt limit~\cite{Schmidt1937}), 
\begin{align}
    \mu_{\He} &= \kappa_0 - \kappa_1 = \mu^{\mathrm{exp}}_n.
\end{align}
In addition, the $\He$ charge, matter, and magnetic point radii are all equal in the Wigner-SU(4) limit, i.e.,
\begin{align}
\label{eq:radii-relations-full-Wigner}
    \langle r^2_{0}\rangle^C_{\He} = \langle r^2_{0}\rangle^\#_{\He} = \langle r^2_{0}\rangle^{v}_{\He} = \langle r^2_{0}\rangle^{s}_{\He}.
\end{align}

This section considers how the Coulomb effect modifies the role of  Wigner-SU(4) symmetry. Specifically, it will be argued that the Schmidt limit for the $\He$ magnetic moment still holds as long as $\delta =0$. 
In contrast, the relation for the radii in Eq.~\eqref{eq:radii-relations-full-Wigner} is partially broken by the Coulomb interaction. It will also be argued that the isoscalar and isovector magnetic point radii are still equal at $\delta = 0$ with nonperturbative Coulomb,
\begin{align}
\label{eq:mag-radii-relations-partial-Wigner}
    \langle r^2_{0}\rangle^{s}_{\He} = \langle r^2_{0}\rangle^{v}_{\He}, 
\end{align}
but the relation between the charge, matter, and magnetic radii becomes
\begin{align}
\label{eq:radii-relations-partial-Wigner}
    \langle r^2_{0}\rangle^{s/v}_{\He} + 2\langle r^2_{0}\rangle^C_{\He} = 3\langle r^2_{0}\rangle^\#_{\He}.
\end{align}

Finally, it will be argued that the inclusion of the Coulomb interaction promotes the Wigner-SU(4) breaking in the $\He$ binding energy from $\mathcal{O}({\delta^2})$ to $\mathcal{O}({\delta})$. 
In contrast, it will be shown that $\mu_{\He}$ is expected to receive a Wigner-SU(4) breaking of $\sim 2\delta^2/9$, regardless of the presence of the Coulomb interaction.
In addition, the analysis on the $\delta$ scaling, together with the numerical results in Sec.~\ref{sec:results}, provides evidence for an estimate of the three-body scale $\kappa^*_3 \approx 80 \sim 100$~MeV, close to the $\triton$ or $\He$ binding momentum as one might naively expect. This corresponds to a Wigner-SU(4) expansion parameter $\delta/\kappa^*_3 \approx30\%$, roughly the same as the \ac{eftnopi} expansion parameter.

\subsection{Partial Wigner-SU(4) limit}
To distinguish the Wigner-SU(4) limit with or without Coulomb, this paper refers to the limit of $\delta \to 0$ in the presence [absence] of the Coulomb interaction as the partial [full] Wigner-SU(4) limit. 
Specifically, in the partial Wigner-SU(4) limit the \ac{pp} propagator  takes the same form as that in the physical limit, while the other two dibaryon propagators take their forms in the full Wigner-SU(4) limit; i.e., in the partial Wigner-SU(4) limit, the dibaryon propagators in Eq.~\eqref{eq:dibaryon_prop} become
\begin{align}
    D_t(E,q) &\to D_{ws}(E,q)\nonumber\\
    D_s(E,q) &\to D_{ws}(E,q)\\\nonumber
    D_{pp}(E,q) &\to D_{pp}(E,q),
\end{align}
where $D_{ws}(E,q)$ is given by
\begin{align}
\label{eq:Dw}
    D_{ws}(E,q) = & \frac{1}{\gamma - \sqrt{\frac{q^2}{4} - M_N E - i\epsilon}}.
\end{align}

To study the impact of partial Wigner-SU(4) symmetry on the $\He$ wavefunction, one can rotate the $\He$ vertex and wavefunction into a three-channel Wigner basis, defined by the transformation matrix
\begin{align}
\label{eq:Qmatrix}
   \Qb = 
   \begin{pmatrix}
       -1/\sqrt{2}& 1/\sqrt{6}&\sqrt{1/3}\\
       1/2& -1/2\sqrt{3}&\sqrt{2/3}\\
        1/2&\sqrt{3}/2& 0
   \end{pmatrix}.
\end{align}
The integral equation for the $\He$ vertex function in Eq.~\eqref{eq:vertex_function} becomes
\begin{align}
    \label{eq:vertex_function_wigner}
    \Gb_w(E, p) =& \Bb_w + \int_0^\Lambda \frac{q^2dq}{2\pi^2}\Kb_w(E,p,q) \Gb_w(E,q),
\end{align}
where,
\begin{align}
    \label{eq:wigner_transform}
     \Gb_w = \Qb \Gb, \qquad \Bb_w = \Qb \Bb, \qquad \Kb_w = \Qb \Kb \Qb^\intercal.
\end{align}
Similarly, the wavefunctions $\widetilde{\boldsymbol{\psi}}_{sc}$ in Eq.~\eqref{eq:psi_sc_he3_tilde} and $\widetilde{\boldsymbol{\psi}}_{c}$ in Eq.~\eqref{eq:psi_c_he3_tilde} can be transformed into the Wigner basis,
\begin{align}
    \label{eq:psi_wigner_transform}
    \widetilde{\boldsymbol{\psi}}_{sc,w} = \Qb \widetilde{\boldsymbol{\psi}}_{sc}, \qquad \widetilde{\boldsymbol{\psi}}_{c,w} = \Qb \widetilde{\boldsymbol{\psi}}_{c}. 
\end{align}

An important fact about $(\Gb_w)_i$, $(\widetilde{\boldsymbol{\psi}}_{sc,w})_i$, and $(\widetilde{\boldsymbol{\psi}}_{c,w})_i$ is that their $i=2$ component vanishes only in the full (but not partial) Wigner-SU(4) limit, while their $i=3$ component  vanishes in both limits. (The subscript $i$ indicates the $i$-th component of the vector in the three-channel formalism.) To see this, first observe that the inhomogeneous term in Eq.~\eqref{eq:vertex_function_wigner} only has its first component being non-zero,  
\begin{align}
\label{eq:Bbw}
    (\Bb_w)_1= -\sqrt{6}, \qquad (\Bb_w)_2= (\Bb_w)_3 = 0.
\end{align}
In addition, the transformation of $\Kb$ in Eq.~\eqref{eq:wigner_transform} suggests that $\Xb$ and $\Db_d$ that enter the kernel in Eq.~\eqref{eq:kernel_strong} transform, respectively, as
\begin{align}
    \label{eq:Xb_w}
    \Xb_w = \Qb \Xb \Qb^\intercal =&
    -\frac{1}{4}\begin{pmatrix}
        -2& 0&0\\
        0& 1&0\\
        0&0&1
    \end{pmatrix},
\end{align}
and
\begin{widetext}
\begin{align}
    \label{eq:Db_wigner_basis}
    \Db_w = \Qb \Db_d \Qb^\intercal 
        =&
        {\renewcommand{\arraystretch}{2.4}%
        \begin{pmatrix}
                \dfrac{3D_t+D_s+2D_{pp}}{6}&\dfrac{-3\sqrt{2}D_t-\sqrt{2}D_s+4\sqrt{2}D_{pp}}{6}&\dfrac{-\sqrt{2}D_t+\sqrt{2}D_s}{6}\\
            \cdots& \dfrac{3D_t+D_s+8D_{pp}}{12}&\dfrac{D_t-D_s}{6}\\
             \cdots &\cdots&\dfrac{D_t+3D_s}{4}
        \end{pmatrix}},
\end{align}
\end{widetext}
where the symmetric entries are omitted. In the full Wigner-SU(4) limit, $\Kb_w$ only involves the strong interaction piece $\Kb_s$, and $\Db_w$ and thus the kernel $\Kb_w$ become diagonal. This means that in the full Wigner-SU(4) limit, all three channels in the Wigner basis decouple and that $(\Gb_w)_i$,  $(\widetilde{\boldsymbol{\psi}}_{sc,w})_i$, and $(\widetilde{\boldsymbol{\psi}}_{c,w})_i$ vanish for $i = 2,3$. To go from the full to the partial Wigner-SU(4) limit, the Coulomb interaction is included with $\delta = 0$ fixed. The kernel $\Kb_w$ receives Coulomb contributions, but it remains block-diagonal with the last channel decoupled from the first two.
This can be verified by transforming all the other relevant \ac{CCS} matrices into the Wigner basis, including $\Nb_c$, $\Xb_\beta$, and $\Xb_\gamma$ defined in Eqs.~\eqref{eq:Nc}, \eqref{eq:Xb_beta}, and \eqref{eq:Xb_gamma}, respectively. 
As a result, $(\Gb_w)_3$,  $(\widetilde{\boldsymbol{\psi}}_{sc,w})_3$, and $(\widetilde{\boldsymbol{\psi}}_{c,w})_3$ remain decoupled from the other two channels and, combined with Eq.~\eqref{eq:Bbw}, they must vanish in the partial Wigner-SU(4) limit.

\subsection{Impact on Helium-3 moments and radii}
\label{sec:wig_impact_on_He3}
The $\He$ form factor in Eq.~\eqref{eq:psi-O-psi-0} can be written in the Wigner basis to study the impact of partial Wigner-SU(4) symmetry on $\He$ observables. This requires the expression of $\Mb^\xi_{\alpha,i}$ in the Wigner basis.  Among the $\He$ charge and magnetic moments,  the isovector magnetic moment is the only one that receives a non-zero Coulomb correction since the charge and spin of $\He$ are not affected by the inclusion of Coulomb. Transforming $\Mb^v_{a,i}$ and $\Mb^v_{b,i}$ in App.~\ref{app:cc_matrices} to the Wigner basis and summing over $i$ gives
\begin{equation}
\label{eq:Mv}
\begin{aligned}
   & \Mb^{v}_{W,a} \equiv\Qb \left(\sum_{i=1}^3\Mb^{v}_{a,i}\right)\Qb^\intercal = \frac{\kappa_1}{2}\begin{pmatrix}
       -1 & 0&0\\
       0& -1&0\\
        0&0& \frac{5}{3}
    \end{pmatrix},\\
    &\Mb^{v}_{W,b} \equiv \Qb \left(\sum_{i=1}^3\Mb^{v}_{b,i}\right)\Qb^\intercal = -\frac{\kappa_1}{2}\begin{pmatrix}
    2 & 0&0\\
    0& -1&0\\
    0&0& \frac{5}{3}
\end{pmatrix},
\end{aligned}
\end{equation}
and similarly for the nucleon number operator,
\begin{equation}
\label{eq:Mnum}
\begin{aligned}
    &\Mb^{\#}_{W, a} \equiv \Qb\left(\sum_{i=1}^3\Mb^{\#}_{a,i} \right)\Qb^\intercal= \frac{3}{2}\begin{pmatrix}
    1 & 0&0\\
    0& 1&0\\
    0&0& 1
\end{pmatrix},\\
&\Mb^{\#}_{W, b} \equiv \Qb\left(\sum_{i=1}^3\Mb^{\#}_{b,i}\right)\Qb^\intercal = -\frac{3}{2}\begin{pmatrix}
    -2 & 0&0\\
    0& 1&0\\
    0&0& 1
\end{pmatrix}.
\end{aligned}
\end{equation}
Comparing Eqs.~\eqref{eq:Mv} and~\eqref{eq:Mnum} shows that 
\begin{equation}
\label{eq:Mv-Mnum-relation}
\begin{aligned}
    \left[\Mb^{v}_{W,a}\right]_{2} &= -\frac{\kappa_1}{3}\left[\Mb^{\#}_{W,a}\right]_{2},\\ \left[\Mb^{v}_{W,b}\right]_{2} &= -\frac{\kappa_1}{3}\left[\Mb^{\#}_{W,b}\right]_{2}.
\end{aligned}
\end{equation}
where $\left[\Mb\right]_{k}$ is the submatrix of $\Mb$ formed by its first $k$ rows and columns. 
It is straightforward to show that the relation in Eq.~\eqref{eq:Mv-Mnum-relation} also holds for type-($d$), ($e$), and ($f$) contractions. 
Combined with the fact that $(\widetilde{\boldsymbol{\psi}}_{sc,w})_3$ and $(\widetilde{\boldsymbol{\psi}}_{c,w})_3$ vanish in the partial Wigner-SU(4) limit, Eq.~\eqref{eq:Mv-Mnum-relation} indicates that the $\He$ matrix element of the isovector magnetic operator at zero momentum transfer is equivalent to that of the nucleon number operator times a factor of $-\kappa_1/3$.
Therefore, the $\He$ magnetic moment in the partial Wigner-SU(4) limit is given by
\begin{align}
\label{eq:mu_He_wigner}
    \mu_{\He} = \kappa_0 - \kappa_1,
\end{align}
which holds regardless of the inclusion of non-perturbative Coulomb interactions. 

To show the relations for the $\He$ point radii in Eqs.~\eqref{eq:mag-radii-relations-partial-Wigner} and~\eqref{eq:radii-relations-partial-Wigner}, first observes that for the derivative of the form factor in Eq.~\eqref{fig:He3formfactor}, the summation over $i$ involves both $\Mb^\xi_{\alpha,i}$ and the derivatives of $\rho_i^{\ell\ell'}$ with respect to $k^2$ at $k=0$, as noted in the discussion below Eq.~\eqref{eq:two-body-rho-derivative}. Therefore,  a relation for $\Mb^\xi_{\alpha,i}$ that holds for each $i$ separately must be identified. It can be verified that
\begin{align}
    \label{eq:Mvi-Msi-relation}
    \frac{1}{\kappa_1}\left[\Mb^v_{W,\alpha,i}\right]_{2} = -\frac{1}{\kappa_0}\left[\Mb^s_{W,\alpha,i}\right]_{2} 
\end{align}
and
\begin{align}
\label{eq:Mall-relation}
\frac{1}{\kappa_0}\left[\Mb^s_{W,\alpha,i}\right]_{2} + \frac{1}{2}\left[2\Mb^C_{W,\alpha,i}\right]_{2} = \frac{1}{3}\left[3\Mb^\#_{W,\alpha,i}\right]_{2},
\end{align}
for all $\alpha$ and $i$, where $\Mb^\xi_{W,\alpha,i}$ is $\Mb^\xi_{\alpha,i}$ in the Wigner basis. Eqs.~\eqref{eq:mag-radii-relations-partial-Wigner} and~\eqref{eq:radii-relations-partial-Wigner} can be obtained using the relations above and the fact that $(\widetilde{\boldsymbol{\psi}}_{sc,w})_3$ and $(\widetilde{\boldsymbol{\psi}}_{c,w})_3$ vanish in the partial Wigner-SU(4) limit.

\subsection{Breaking of the Full and Partial Wigner-SU(4) symmetry}
\label{sec:Wig-break}

This section analyzes the sizes of $\delta$-correction to the $\He$ observables, particularly the $\He$ binding energy and magnetic moment, when expanding around the full and partial Wigner-SU(4) limit. 
As argued below Eq.~\eqref{eq:Db_wigner_basis}, $(\Gb_w)_i$, $(\widetilde{\boldsymbol{\psi}}_{sc, w})_i$, and $(\widetilde{\boldsymbol{\psi}}_{c, w})_i$ must vanish for $i=2,3$ in the full Wigner-SU(4) limit. Their perturbative correction in $\delta$ is driven by the $\delta$-expansion of $\Db_w$ in Eq.~\eqref{eq:Db_wigner_basis}. In particular, the top-left component of $\Db_w$ receives no $\mathcal{O}(\delta)$ correction without the Coulomb interaction because
\begin{align}
\label{eq: Db_w_11_Wigner_expansion_no_Coulomb}
   (\Db_w)_{11} =&\frac{3D_t + D_s + 2D_{pp}}{6}  \\
 &\hspace{-25pt}\stackrel{ (\textrm{no Coulomb}) }{\longrightarrow} \frac{D_t + D_s}{2} \quad  \\
     =& D_{ws}\sum_{n =0}^{\infty} \left(\delta D_{ws}\right)^{2n},
\end{align}
where $(\Db_w)_{ij}$ is the $i,j$-th component of $\Db_w$ in the three-channel formalism. $D_{ws}$ is the dibaryon propagator in the Wigner-SU(4) limit defined in Eq.~\eqref{eq:Dw}. A similar expansion on the off-diagonal elements of $\Db_w$ starts at $\mathcal{O}(\delta)$. Therefore, the $\delta$-correction to $(\Gb_w)_i$ and the $\He$ wavefunctions starts at $\mathcal{O}(\delta^2)$ for $i=1$ and at $\mathcal{O}(\delta)$ for $i=2,3$. This means that in the absence of the Coulomb interaction, the Wigner-SU(4) breaking on the $\He$ binding energy (or equivalently the $\triton$ binding energy) starts at $\mathcal{O}(\delta^2)$.\footnote{An explicit expansion of the binding energy shift in the two-channel formalism can be found in Ref.~\cite{Vanasse:2016umz}.}

In contrast, any correction to the first two components of $\Gb_w$ or the wavefunctions has no impact on $\mu_{\He}$ as shown previously. In other words, the $\delta$ correction to the $\mu_{\He}$ comes from the $\delta$ correction to $(\Gb_w)_3$.  Using the expansion of $(\Db_w)_{13}$ that connects $(\Gb_w)_3$ and $(\Gb_w)_1$,
\begin{align}
(\Db_w)_{13} &=\frac{-\sqrt{2}D_t + \sqrt{2}D_s}{6} \\
&=\frac{\sqrt{2}}{3}D_{ws}\sum_{n=0}^\infty \left(\delta D_{ws}\right)^{2n+1},
\end{align}
it can be shown that the leading non-vanishing $\delta$ correction to $\mu_{\He}$ is expected to be of order $\sim \delta^2$; more precisely the correction is $\sim {2\delta^2}/{9}$ as leaving  out the factor of $2/9$ could obscure the $\delta$ power counting for $\delta/\kappa^*_3 \sim30\%$. This factor reflects the fact that only a small portion of the $\delta$ correction to the $\He$ wavefunction actually contributes to that of $\mu_{\He}$.

The $\delta$ scaling becomes different in the presence of Coulomb, which can be observed from the fact that the $\delta$ correction to $(\Db_w)_{11}$ starts at $\mathcal{O}(\delta)$. An important consequence is that $(\Gb_w)_1$ now obtains an $\mathcal{O}(\delta)$ correction. This in turn promotes the $\delta$ correction to the $\He$ binding energy from $\mathcal{O}(\delta^2)$ to $\mathcal{O}(\delta)$ when expanding around the partial (as opposed to the full) Wigner-SU(4) limit.
On the other hand, the $\delta$ correction to $\mu_{\He}$ receives no such promotion because the $\delta$ scaling of $(\Db_w)_{13}$, and thus that of  $(\Gb_w)_3$, is unaffected by Coulomb. Consequently, the $\delta$ correction to $\mu_{\He}$ still starts at  $\sim2\delta^2/9$.

\section{Results and Discussion}
\label{sec:results}
\subsection{Coulomb corrections to Helium-3 binding energy and magnetic moment}

\begin{figure}[htb!]
    \centering
    \includegraphics[width=0.9\columnwidth]{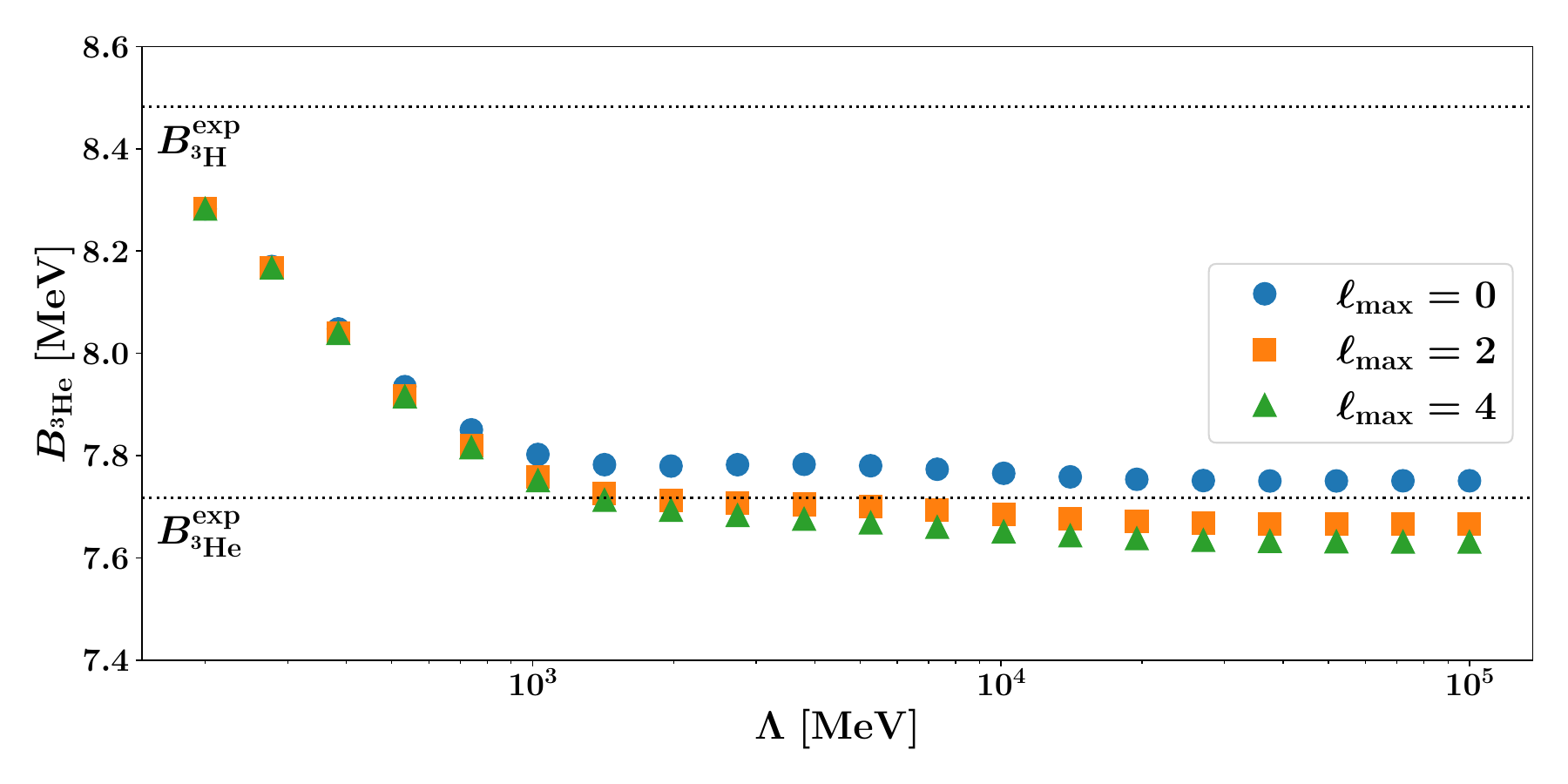}
    \\
    \includegraphics[width=0.95\columnwidth]{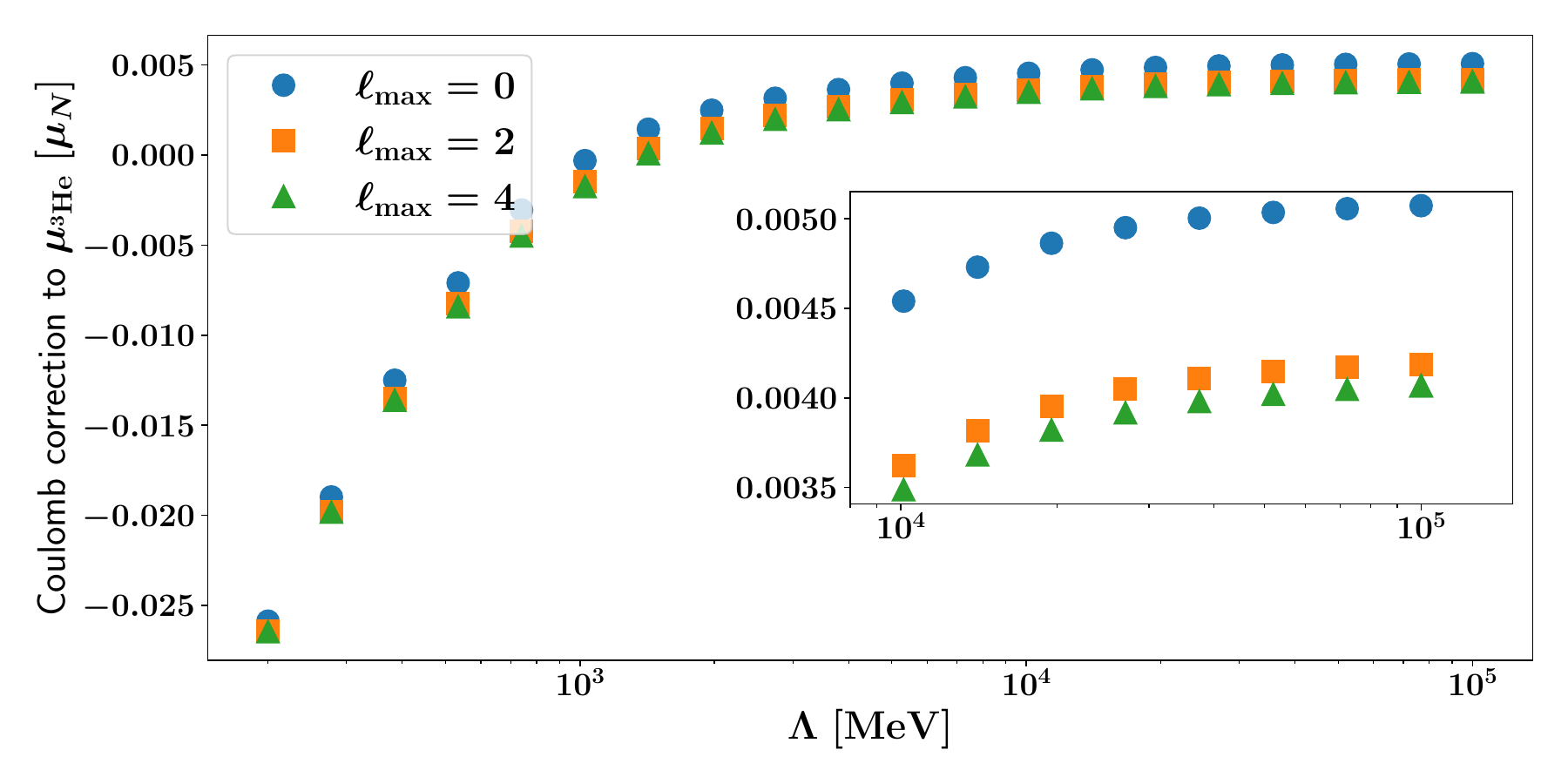}
    \\
    \caption{ Cutoff dependence of $B_{\He}$ (top). Cutoff dependence of the Coulomb correction to $\mu_{\He}$  (bottom). }
    \label{fig:B3He_convg}
\end{figure}

Figure~\ref{fig:B3He_convg} shows $B_{\He}$ and the Coulomb correction to $\mu_{\He}$ as a function of the sharp momentum cutoff, $\Lambda$, and angular momentum cutoff, $\ell_{\mathrm{max}}$. The results are obtained using a quadratic extrapolation of $m_\gamma \to 0$. Comparing the results with $\ell_{\mathrm{max}} = 0, 2, 4$ demonstrates a clear convergence with respect to $\ell_{\mathrm{max}}$. The relative uncertainty resulting from a truncation at $\ell_{\mathrm{max}} = 4$ can be estimated using the difference between the results with $\ell_{\mathrm{max}}=2$ and $4$; this is $\lesssim 5\%$ of the Coulomb correction and will be shown in parenthesis below for all relevant observables. 
At $\ell_\mathrm{max} = 4$ and $\Lambda = 10^5$~MeV, $B_{\He}$ is found to be $7.63(3)$~MeV in the $m_\gamma \to 0$ limit. This agrees well with the previous \ac{eftnopi} calculations of $B_{\He} = 7.66$~MeV in Ref.~\cite{Ando:2010wq} and $B_{\He} \approx 7.61$~MeV extracted from Ref.~\cite{Konig:2014ufa} at their largest cutoff. 
To compare with experiments, the splitting between $B_{\triton}$ and $B_{\He}$ serves as a better metric than $B_{\He}$ itself as the \ac{LO} three-body force is fit to reproduce $B^{\mathrm{exp}}_{\triton} = 8.48$~MeV. The splitting obtained from this work is $0.85(3)$~MeV, which is in good agreement with the experimental value of $0.76$~MeV with a relative difference of $\sim 15\%$, well within the naive \ac{LO} \ac{eftnopi} error of $\sim  30\%$.

The Coulomb correction to $\mu_{\He}$ is found to be $0.0041(1) \mu_N$ at $\ell_{\mathrm{max}}=4$ and $\Lambda=10^5$~MeV. This value is only $0.2\%$ of the LO $\mu_{\He}$ of $1.868\mu_N$ without the Coulomb interaction, and is close to the expected size of N$^5$LO terms in the \ac{eftnopi} expansion. In fact, it differs from the naively estimated Coulomb correction of $\sim8\%$ by a factor of 40. This can be understood using Wigner-SU(4) symmetry and will be discussed in the next section. Including the Coulomb correction, the \ac{LO} \ac{eftnopi} result of $\mu_{\He} = -1.864(1) \mu_N$ underpredicts the experimental value of $\mu^{\mathrm{exp}}_{\He} = -2.127\mu_N$ by $ 15\%$, within the naive \ac{LO} \ac{eftnopi} error of $\sim  30\%$.

\subsection{Wigner-SU(4) breaking in Helium-3 binding energy and magnetic moment}

This section analyzes the dependence of $B_{\He}$ and $\mu_{\He}$ on the Wigner-SU(4) breaking parameter $\delta$ and then uses them to extract the dimensionless Wigner-SU(4) expansion parameter, $\delta/\kappa_3^*$. The results shown below are obtained using a three-body force fixed to the physical value of $B^{\mathrm{exp}}_{\triton} = 8.48$~MeV; i.e., the three-body force does not change with $\delta$. 

\begin{figure}[htb!]
    \centering
    \includegraphics[width=0.9\columnwidth]{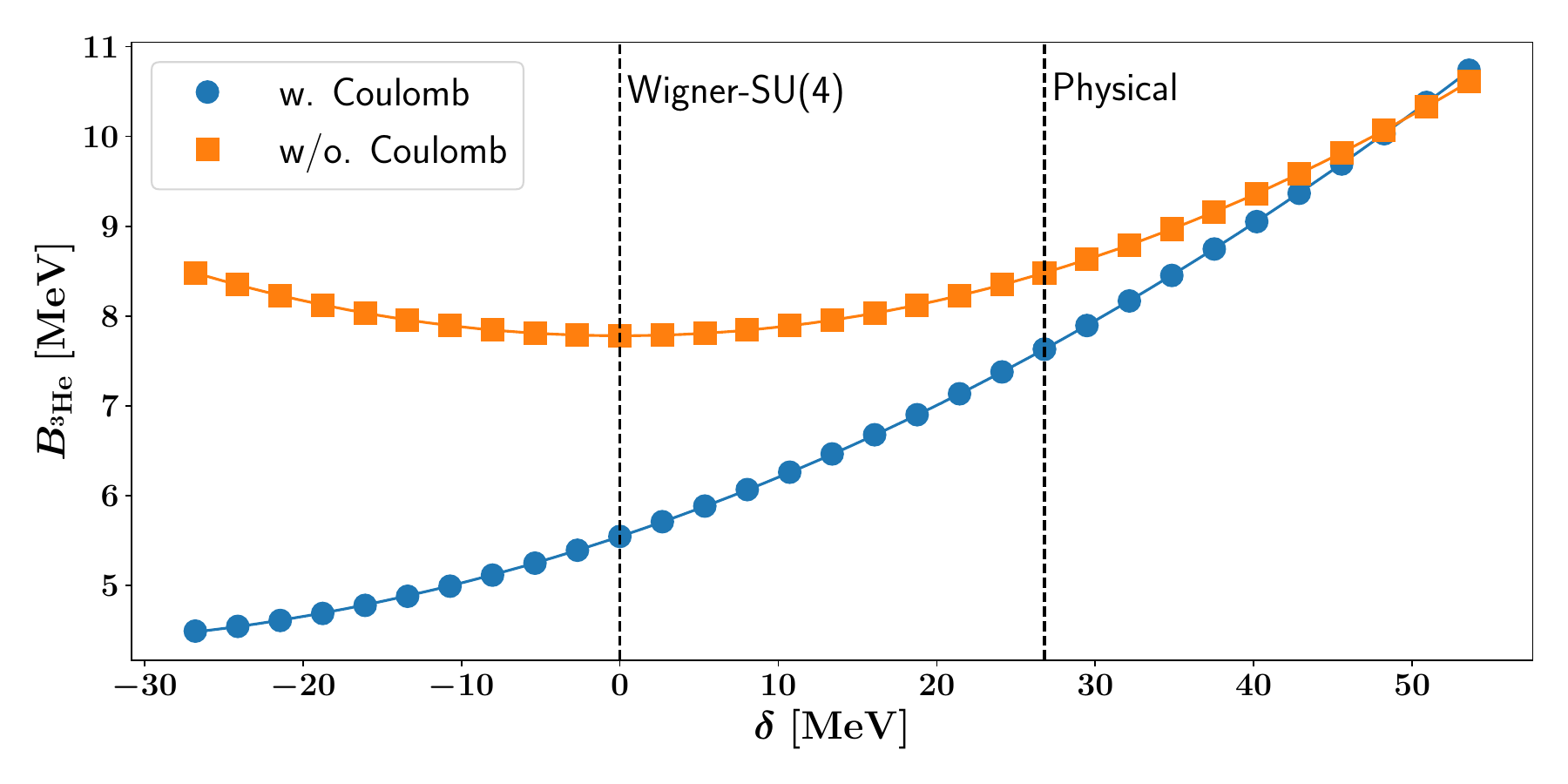}
     \includegraphics[width=0.9\columnwidth]{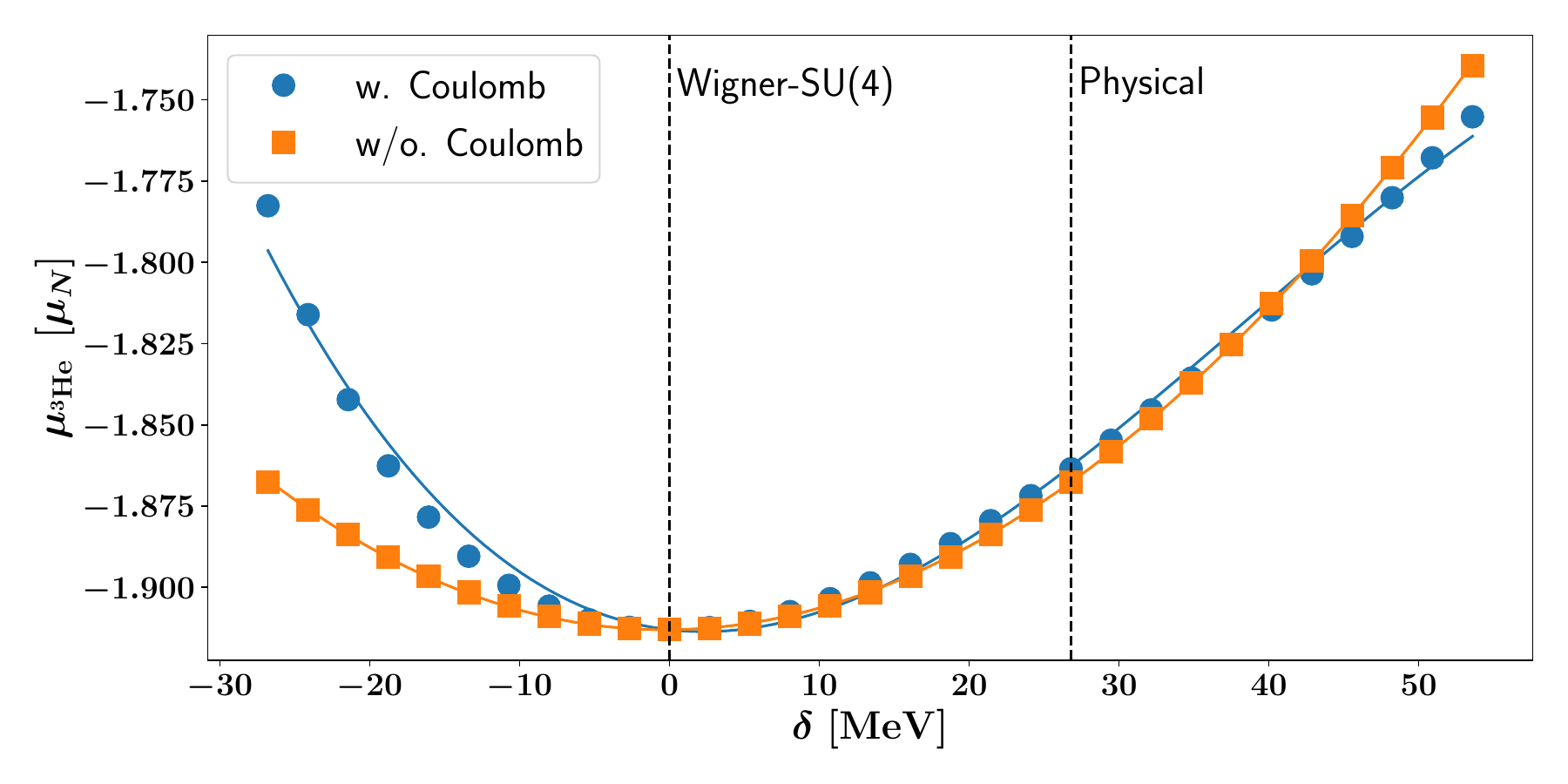}
    \caption{The $\delta$-dependence of $B_{\He}$ (top) and $\mu_{\He}$ (bottom) with and without the Coulomb interaction. The curves are the best fits to the polynomials in Eq.~\eqref{eq:poly_delta_fit} with the fitted coefficients shown in Table~\ref{tab:Wigner_fit_coefficients}. Vertical lines indicate the location of the full/partial Wigner-SU(4) limit ($\delta=0$) and the physical value of $\delta$.}
    \label{fig:3N_binding_delta}
\end{figure}
Fig.~\ref{fig:3N_binding_delta} shows the $\delta$ dependence of $B_{\He}$ and $\mu_{\He}$ with (w.) and without (w/o.) non-perturbative Coulomb interaction. The $\delta$ correction to $B_{\He}$ only contains even powers of $\delta$ without Coulomb, while that with Coulomb contains both even and odd orders of $\delta$. This applies similarly to $\mu_{\He}$, except for the lack of an $\mathcal{O}(\delta)$ term in the case without Coulomb. 
A polynomial of the form
\begin{align}
\label{eq:poly_delta_fit}
    f(\delta) &= A\left(1 + a_1 \delta + a_2 \delta^2 + a_3\delta^3\right) 
\end{align}
is fit to the results and the coefficients are shown in Table~\ref{tab:Wigner_fit_coefficients}. Note that Eq.~\eqref{eq:poly_delta_fit} makes no assumption on the evenness or oddness of the $\delta$ scaling. In general, the $\delta$ dependence of the results are well captured by the fitted curves.

As can be observed in Fig.~\ref{fig:3N_binding_delta}, the splitting between the results with and without Coulomb are much greater at a negative value of $\delta$ than a positive $\delta$. This is expected because,  as $\delta$ moves too far away from its physical value, the Coulomb effect that distinguishes the \ac{pp} channel from the ${}^1S_0 (np)$ channel becomes more significant implicitly, even though the Coulomb $T$-matrix itself is unchanged.
Another important observation from the top panel of Fig.~\ref{fig:3N_binding_delta} is that $B_{\He}$ with Coulomb at the physical value of $\delta$ is close to that without Coulomb at $\delta = 0$. This suggests a significant cancellation between the Coulomb and $\delta$ corrections to $B_{\He}$ at the physical value of $\delta$. 
As $\delta$ grows to $\approx 50$ MeV, which is roughly twice its physical value, the results of $B_{\He}$ with and without Coulomb cross each other, indicating a sign change of the Coulomb correction. A similar crossing is also observed in $\mu_{\He}$ at $\delta \approx 40$~MeV on the lower panel of Fig.~\ref{fig:3N_binding_delta}. Since $\mu_{\He}$ is fixed by the Schmidt limit of $\kappa_0 - \kappa_1$ at $\delta = 0$ both with and without Coulomb and no abrupt change is expected, the Coulomb correction to $\mu_{\He}$ between $\delta = 0$ and $\approx 40$~MeV appears much smaller than that to the binding energy. This is consistent with the $\sim \delta$ vs. $\sim 2\delta^2/9$ suppression of the Coulomb correction to $B_{\He}$ and $\mu_{\He}$, respectively, as discussed in Sec.~\ref{sec:Wig-break}.

\begin{table}
    \centering
    \caption{Coefficients obtained by fitting Eq.~\eqref{eq:poly_delta_fit} to the $\delta$-dependence of $B_{\He}$ and $\mu_{\He}$ in Fig.~\ref{fig:3N_binding_delta}. Starred values are fixed to the Schmidt limit of $\mu_{\He} = \kappa_0 - \kappa_1$.}
    \label{tab:Wigner_fit_coefficients}
    \begin{tabular}{c|c|cccc}
         & & \makecell{$A$\\ $[\mathrm{MeV}]/[\mu_N]$} & \makecell{$a_1$\\$(\times 10^2)$} &  \makecell{$a_2$\\$(\times 10^4)$} & \makecell{$a_3$\\$(\times 10^6)$}\\
         \hline
         \multirow{2}{*}{\makecell{w.\\ Coulomb}} & $B_{\He}$ & 5.54 & 1.06 & $1.29$ & -0.02\\
        & $\mu_{\He}$ & -1.913* & 0.03 & -0.61 & 0.52\\
        \hline
        \multirow{2}{*}{\makecell{w/o.\\ Coulomb}} & $B_{\He}$ & 7.78 & -0.002 & 1.25 & 0.003\\
        & $\mu_{\He}$ & -1.913*  &-0.002& -0.33 &0.04\\
        \hline\hline
    \end{tabular}
\end{table}
The fitted coefficients in Table~\ref{tab:Wigner_fit_coefficients} can be used to estimate the dimensionless expansion parameter $\delta/\kappa^*_3$ under the assumption of naturalness, i.e., dimensionless coefficients should be all be order one.  Table~\ref{tab:Wigner_fit_coefficients} shows a clear order-of-magnitude separation among the fitted coefficients. The small coefficients ($\leq 0.03$) are considered to reflect the evenness and oddness of the $\delta$ scaling and thus not used to estimate $\delta/\kappa^*_3$. In addition, for coefficients of $\mu_{\He}$, the factor of  $2/9$ in the $\sim 2\delta^2/9$ suppression on $\mu_{\He}$ should be accounted for. Finally, taking ratios of the fitted coefficients for each observable yields a $\kappa^*_3$ of roughly $80 \sim 100$~MeV, close to the $\triton$ or $\He$ binding momentum as one naively expects. With $\delta \approx 27$~MeV in the physical world, the Wigner-SU(4) expansion parameter $\delta/\kappa^*_3$ is approximately $30\%$, roughly the same as the \ac{eftnopi} expansion parameter. In particualr,  this correponds to a suppression of $2\delta^2/9\kappa^{*2}_3 \approx 3\% $ on the $8\%$ Coulomb correction to $\mu_{\He}$, in agreement with the small Coulomb correction of only $0.2\%$ shown in the previous section.

\subsection{Helium-3 radii and Wigner-SU(4) limits}
\begin{figure}[htb!]
    \centering
    \includegraphics[width=0.95\columnwidth]{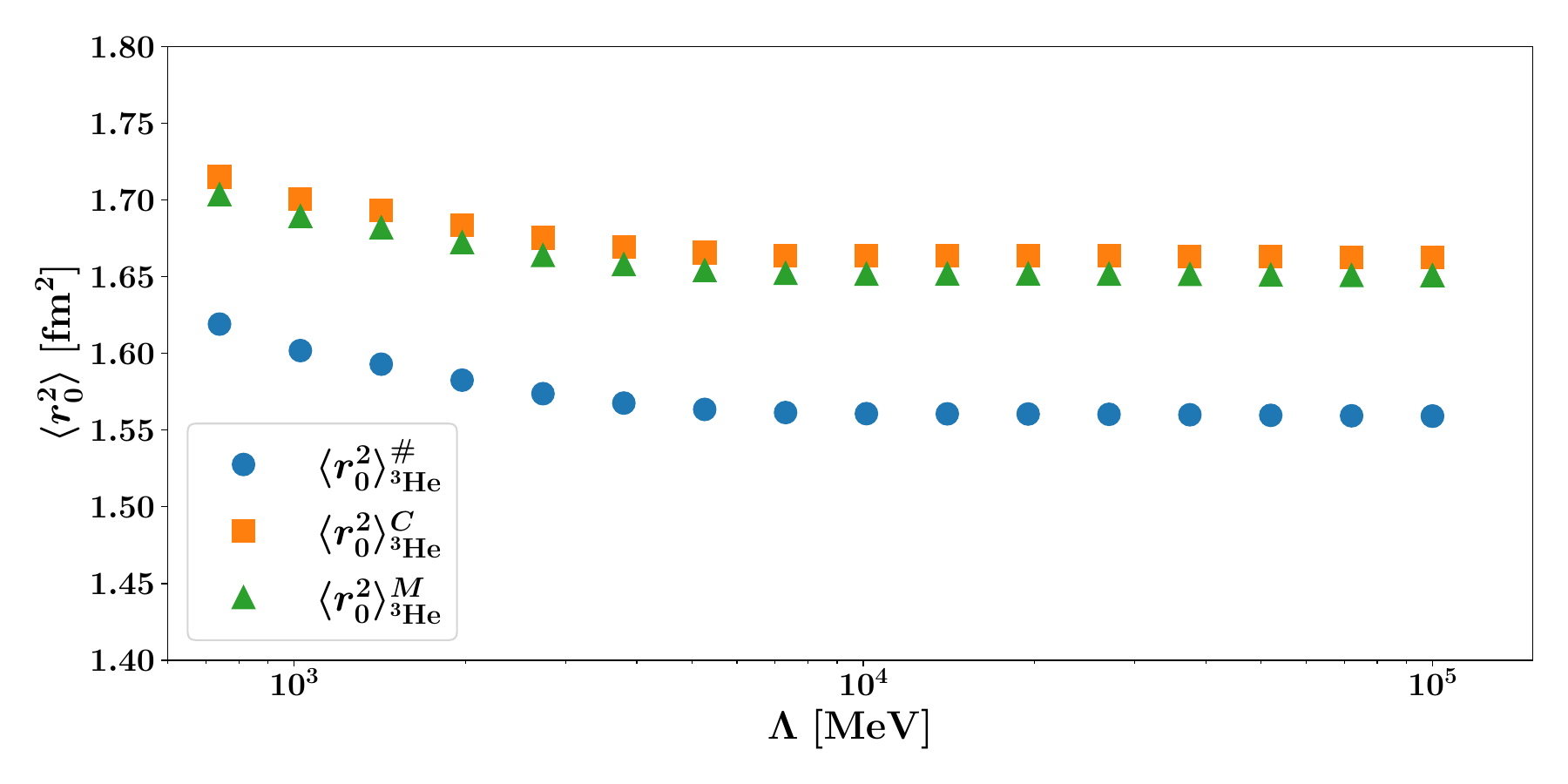}
    \caption{Cutoff dependence of the $\He$ matter, charge, and magnetic point radii squared with at $\ell_{\mathrm{max}} = 4$. Results at $\Lambda = 10^5$~MeV correspond to those on the row of ``w. Coulomb'' and ``Phys'' in Table~\ref{tab:radii}.}
    \label{fig:He3_radii_convg}
\end{figure}

\begin{table*}[t]
    \centering
    z\caption{$\He$ point radii at \ac{LO} in \ac{eftnopi} with (w.) and without (w/o.) the Coulomb interaction and in the physical (Phys.), full Wigner-SU(4) (Wig), or partial Wigner-SU(4) limit  (PWig). 
    The full $\He$ magnetic radius $\langle r\rangle^{M}_{\He}$ is also presented. 
    The $\He$ binding energy and magnetic moment are also shown for completeness.
    Uncertainties from the $\ell_{\mathrm{max}} = 4$ trunctation are shown; for  $\langle r\rangle^{M}_{\He}$ the uncertainty also includes the experimental ones propagated through Eq.~\eqref{eq:full_mag_radius}. An \ac{LO} \ac{eftnopi} truncation error of $\sim 15\%$ [$30\%$] applies to the radii [$B_{\He}$ and $\mu_{\He}$] and is not shown explicitly.
    Experimental value for  $\langle r_{0}\rangle^{C}_{\He}$ is calculated using Eq.~\eqref{eq:point_charge_radius}. The starred value is used to fit the three-body force.}
    \label{tab:radii}
    \begin{tabular}{c|c|cccccc}
         &  & \makecell{$\langle r_{0}\rangle^\#_{\He}$ \\ $[\text{fm}]$} &\makecell{$\langle r_{0}\rangle^C_{\He}$ \\ $[\text{fm}]$} & \makecell{$\langle r_{0}\rangle^M_{\He}$ \\ $[\text{fm}]$} & \makecell{$\langle r\rangle^M_{\He}$ \\ $[\text{fm}]$}& \makecell{$B_{\He}$ \\  $[\text{MeV}]$ } &  \makecell{$\mu_{\He}$ \\  $[\mu_N]$ } \\
         \hline
        \multirow{2}{*}{\makecell{w.\\ Coulomb}} & Phys.& 1.249(2) & 1.289(2)  & 1.285(2) & 1.549(5)  &7.63(3)  & -1.864(1) \\
        & PWig &1.326(3)&  1.352(3) & 1.273(3)& 1.539(5)& 5.54(3)&-1.913\\
         \hline
        \multirow{2}{*}{\makecell{w/o.\\ Coulomb}} & Phys. & 1.209& 1.246 & 1.242 &1.513 & 8.48* & -1.868 \\
        & Wig & 1.198 & 1.198&1.198 &1.477& 7.78&-1.913\\
        \hline
        Exp. & & - & 1.7977(11)  & - &  1.965(154)~\cite{Sick:2001rh}&7.718& -2.127\\
        \hline \hline
    \end{tabular}
\end{table*}
The \ac{LO} \ac{eftnopi} results of the $\He$ radii squared  are shown in Figure~\ref{fig:He3_radii_convg} as a function of $\Lambda$.
Results in different limits are summarized in Tabel.~\ref{tab:radii}, where the numbers are obtained at $\Lambda = 10^5$~MeV; for results with Coulomb, an angular momentum cutoff of $\ell_{\mathrm{max}} = 4$ and a quadratic extrapolation to $m_\gamma\to 0$ are also used.
In particular, the full $\He$ magnetic radius $\langle r\rangle^{M}_{\He}$ is obtained by including contributions from the proton and neutron magnetic radii~\cite{Vanasse:2017kgh}
\begin{align}
\label{eq:full_mag_radius}
    \langle r^2\rangle^{M}_{\He} = \langle r^2_{0}\rangle^{M}_{\He} &+ \frac{1}{2}\frac{\mu^{\mathrm{exp}}_p}{\mu_{\He}}\left(1 +\frac{\mu_{\He}^{v}}{\kappa_1}\right)\langle r^2\rangle^{M}_{p} \nonumber \\
   & +  \frac{1}{2}\frac{\mu^{\mathrm{exp}}_n}{\mu_{\He}}\left(1 - \frac{\mu_{\He}^{v}}{\kappa_1}\right)\langle r^2\rangle^{M}_{n},
\end{align}
where $\sqrt{\langle r^2\rangle^{M}_{p}} = 0.851(26)$~fm~\cite{Lee:2015jqa,ParticleDataGroup:2024cfk} and $\sqrt{\langle r^2\rangle^{M}_{n}} = 0.864^{+0.009}_{-0.008}$~fm~\cite{ParticleDataGroup:2024cfk,Belushkin:2006qa,Epstein:2014zua} are the experimental proton and neutron magnetic radii, respectively. $\mu^\mathrm{exp}_{p}$ and $\mu^\mathrm{exp}_{N}$ are the experimental proton and neutron magnetic moments, respectively, given by
\begin{align}
    \mu^\mathrm{exp}_{p} =& \kappa_0 + \kappa_1, \qquad \mu^\mathrm{exp}_{n} = \kappa_0 - \kappa_1.
\end{align}
The $\He$ isovector magnetic moment $\mu_{\He}^{v}$ in Eq.~\eqref{eq:full_mag_radius} has a value of $-2.31\mu_N$ [$-2.30\mu_N$] with [without] the Coulomb interaction and at \ac{LO} in \ac{eftnopi} is related to $\mu_{\He}$ by
\begin{align}
    \mu_{\He}^{v} &= \mu_{\He} - \kappa_0.
\end{align}
In addition, following Ref.~\cite{Vanasse:2017kgh}, the experimental value of the point $\He$ charge radius in Table~\ref{tab:radii} is obtained by subtracting the proton and nucleon radii from the full $\He$ charge radius,
\begin{align}
\label{eq:point_charge_radius}
    \langle r^2_0\rangle^{C}_{\He} = \langle r^2\rangle^{C}_{\He} - \langle r^2\rangle^{C}_{p} - \frac{1}{2}\langle r^2\rangle^{C}_{n}.
\end{align}
where $\sqrt{\langle r^2\rangle^{C}_{p}} = 0.8409\pm 0.0004$~fm~\cite{Xiong:2019umf,Bezginov:2019mdi,ParticleDataGroup:2024cfk} and $\langle r^2\rangle^{C}_{n} = -0.1155\pm 0.0017$~fm$^2$~\cite{ParticleDataGroup:2024cfk}. Using the experimental value of $\langle r^2\rangle^{C}_{\He} = 1.97007(94)$~\cite{CREMA:2025zpo}, this yields $\langle r_0\rangle^{C}_{\He} = 1.7977(11)$.

Comparing the results with and without Coulomb in Table~\ref{tab:radii} shows that the Coulomb interaction increases the $\He$ radii by $\sim 0.04$~fm, i.e., $\sim 4\%$, in the physical limit. This agrees well with the naive estimate of the Coulomb correction of $\alpha M_N/2\sqrt{M_N B^{\mathrm{exp}}_{\He}}\sim 4\%$, where the factor of $1/2$ comes from propagating the uncertainty from the radii squared to radii. 
This means that, unlike the small Coulomb correction to $\mu_{\He}$, the correction to the $\He$ radii is indeed of the same size as an \ac{NNLO} term in the \ac{eftnopi} expansion and must be taken into account to obtain an accurate prediction at \ac{NNLO} or beyond. 
In particular, by neglecting Coulomb contributions Ref.~\cite{Vanasse:2017kgh} obtained an \ac{NNLO} \ac{eftnopi} prediction of $\langle r_{0}\rangle^C_{\He} = 1.74(4)~$fm (with \ac{eftnopi} truncation error shown in parenthesis). 
Including the Coulomb correction calculated in this work would bring the central value to $1.78$~fm, in good agreement with the experimental value of $1.7977(11)$~fm. While similar observations on $\langle r_{0}\rangle^M_{\He}$ can be made by adding its Coulomb correction obtained in this work to its \ac{NLO} prediction in Ref.~\cite{Vanasse:2017kgh}, a detailed analysis of $\langle r_{0}\rangle^M_{\He}$ requires  an \ac{NNLO} study of $\langle r_{0}\rangle^M_{\He}$ and its dependence on two-nucleon magnetic currents (see Ref.~\cite{Vanasse:2017kgh} for this dependence at \ac{NLO}).

It is also noticeable from Table~\ref{tab:radii} that the \ac{LO} \ac{eftnopi} results for the $\He$ radii underpredicts the experimental values by $\sim 35\%$, whether or not the Coulomb interaction is included. This gap is significantly narrowed by the \ac{NLO} \ac{eftnopi} correction, even though it is greater than the naive \ac{LO} \ac{eftnopi} uncertainty of $\sim 15\%$ for the radii~\cite{Vanasse:2015fph,Vanasse:2017kgh}. In fact, the \ac{LO} radii of $\He$ (and $\triton$) can be compared to the universal value in the unitary limit~\cite{Braaten:2004rn} 
\begin{align}
    \langle r_0\rangle^{\mathrm{univ}} = \sqrt{\frac{1+s_0^2}{9M_N B_3}},\qquad  s_0 \approx 1.00624.
\end{align}
Taking $B_3 = B^{\mathrm{exp}}_{\triton}$ [$ B^{\mathrm{exp}}_{\He}$] gives $\langle r_0\rangle^{\mathrm{univ}} = 1.05$~fm [1.10~fm], which is close to the \ac{LO} \ac{eftnopi} results in Table~\ref{tab:radii}. 
In addition, comparing the results in the physical and Wigner-SU(4) limit in Table~\ref{tab:radii} shows that the Wigner-SU(4) limit provides a good approximation.  Despite a Wigner-SU(4) breaking of $\delta/\kappa_3^*\sim 30\%$ in the physical world, the difference between the results in the physical case and in the Wigner-SU(4) limit is only $\approx 4\%$ for the radii, whether or not the Coulomb interaction is included. This provides additional numerical evidence for the utility of the Wigner-SU(4) expansion in few-nucleon systems.

\section{Summary and Outlook}
\label{sec:conclusion}
This work  studies the Coulomb effects and the impact of Wigner-SU(4) symmetry on $\He$ static properties in \ac{eftnopi}. The Coulomb interaction is treated non-perturbatively by including the full off-shell Coulomb $T$-matrix in the three-channel formalism. 
To summarize the \ac{LO} \ac{eftnopi} results, this work obtains a $\He$-$\triton$ energy splitting of $0.85(3)$~MeV, where the indicated error comes from the truncation on partial waves and the result is subject to a $30\%$ \ac{LO} \ac{eftnopi} truncation error (same for all results below). 
This agrees with the experimental value of $0.76$~MeV within the \ac{eftnopi} truncation error, as well as the previous \ac{eftnopi} calculations by Refs.~\cite{Ando:2010wq,Vanasse:2014kxa,Konig:2014ufa,Konig:2015aka}. In addition, the Coulomb correction increases the $\He$ point charge radius by $0.043(2)$~fm and the full magnetic radius by $0.036(2)$~fm. This corresponds to a $\approx 4\%$ correction to their values withut Coulomb of $1.246$~fm and $1.513$~fm, respectively. In contrast, the Coulomb correction to the $\He$ magnetic radius $\mu_{\He}$ is found to be only $-0.0041(1)\mu_N$, which is only $\approx0.2\%$ of the $\mu_{\He} = -1.868\mu_N$ without the Coulomb interaction.  An \ac{LO} \ac{eftnopi} truncation error of $30\%$ applies to all the results above.

These numerical results show a clear hierarchy of Coulomb effects on $\He$ static observables, with a $4\%$ [$10\%$] correction to the $\He$ radii [binding energy] but a much smaller one of $0.2\%$ to the magnetic moment. The correction to the $\He$ radii is of the same size as an \ac{NNLO} term in the \ac{eftnopi} expansion, and thus should be included in order to obtain an accurate prediction at \ac{NNLO} or beyond. In contrast, the Coulomb correction to $\mu_{\He}$ is much smaller than the naive estimate of $ \alpha M_N/\sqrt{M_N B^{\textrm{exp}}_{\He}} \approx 8\%$
and is comparable to a N$^5$LO term in the \ac{eftnopi} expansion. This hierarchy is explained in this work by the Wigner-SU(4) suppression on the Coulomb correction to $\mu_{\He}$; in particular, the Coulomb correction to $\mu_{\He}$ vanishes in the (partial) Wigner-SU(4) limit. In addition, this work also discusses quantitatively the scaling of the $\He$ binding energy and magnetic moment with respect to the Wigner-SU(4) breaking parameter both with and without the Coulomb interaction. Numerical evidence suggests a dimensionless Wigner-SU(4) expansion parameter $\delta/\kappa^*_3 \sim 30\%$. This corresponds to a three-body scale of $\kappa^*_3 \approx 80\sim 100$~MeV, close to the $\triton$ or $\He$ binding momentum as one naively expects.  

Natural next steps include extending the present analysis to higher orders in the \ac{eftnopi} expansion and to other few-nucleon observables. An \ac{NLO} calculation of the $\He$ radii, for example, can be performed by including the \ac{NLO} three-nucleon vertex function (along with other terms) in the calculation of the $\He$ wavefunction. Notably, Refs.~\cite{Vanasse:2014kxa,Konig:2014ufa,Kirscher:2015zoa} have shown that a new isospin-breaking three-body force is needed at \ac{NLO} in the presence of the Coulomb interaction to maintain the \ac{RG} invariance. Alternatively,  the Coulomb interactions in $\He$ observables could be treated fully perturbatively~\cite{Konig:2014ufa,Konig:2015aka,Konig:2016iny,HaAndJared2026}, e.g., as an \ac{NNLO} term in \ac{eftnopi}. For $\He$ this avoids the need for the isospin-breaking three-body force until N$^3$LO, but requires a careful analysis of the perturbative expansion and the associated numerical challenges.

In addition, the formalism in this work for studying the $\He$ form factors can be used to investigate the low-energy $pd$ capture and the impact of Wigner-SU(4) symmetry therein. An \ac{eftnopi} study of  $pd$ capture can provide model-independent predictions with a systematic error that help better understand the existing experimental data~\cite{Cavanna:2018mdc,Stockel:2024hde} and potential model results~\cite{Marcucci:2015yla}. Moreover, Refs.~\cite{Lin:2022yaf,Lin:2024bor} have shown that Wigner-SU(4) symmetry plays an important role in the power counting and error analysis in  $nd$ capture, and it would be interesting to see how Wigner-SU(4) symmetry impacts $pd$ capture as the isospin mirror process of $nd$ capture but with Coulomb. 

\acknowledgments{The author thanks Sebastian Konig, Ha Nguyen, Roxanne Springer, and Jared Vanasse for useful discussions. The author is supported by the U.S. Department of Energy, Office of Science, Office of Nuclear Physics, under Award Number DE-SC0024622, and  by the National Science Foundation under Grant PHY-2044632.}

\onecolumngrid
\appendix

\section{Cluster configuration matrices}
\label{app:cc_matrices}

\subsection{Three-nucleon kernel}
The \ac{CCS} matrix $\Xb$ in Eq.~\eqref{eq:Xb} is obtained from
\begin{align}
\label{eq:Xb_app}
    \Xb &=  \sum_{A,B}
    \begin{pmatrix}
    \frac{\sigma_I}{\sqrt{3}} &  & \\
    & \tau_A\delta_{A3} & \\
    &&\tau_A\delta_{A\pm}
    \end{pmatrix} 
    \begin{pmatrix}
        (P^{\intercal}_J )^\dagger  \\
        (\Bar{P}^{\intercal}_B)^\dagger  \\
        (\Bar{P}^{\intercal}_B)^\dagger
    \end{pmatrix}
    \begin{pmatrix}
        P_I &\Bar{P}_A&  \Bar{P}_A 
    \end{pmatrix}
        \begin{pmatrix}
    \frac{\sigma_J}{\sqrt{3}} &  & \\
    & \tau_B\delta_{B3} & \\
    &&\tau_B\delta_{B\mp}
    \end{pmatrix} \nonumber\\
    &=-\frac{1}{8}
    \begin{pmatrix}
        -1& \sqrt{3}&\sqrt{6}\\
        \sqrt{3}& 1&-\sqrt{2}\\
        \sqrt{6}&-\sqrt{2}& 0
    \end{pmatrix}
\end{align}
where $\sum_A\tau_A\delta_{A\pm} = (\tau_{1} \pm i\tau_{2})/\sqrt{2}$.  The spin and isospin indices of Pauli matrices are suppressed.
The $3\times 3$ matrices on the first row are the projector $\mathbf{P} $in Eq.~\eqref{eq:3Ndoublet-projector} and
selects the spin-doublet channel where three-nucleon bound states live.
The inhomogeneous term $\Bb$ in Eq.~\eqref{eq:3B_vertex_cc_space} comes from
\begin{align}
    \begin{pmatrix}
        \sigma_I\\
        -\tau_A\delta_{A3}\\
        -\tau_A\delta_{A\pm}\sqrt{2}
    \end{pmatrix} =\begin{pmatrix}
    \frac{\sigma_I}{\sqrt{3}} &  & \\
    & \tau_A\delta_{A3} & \\
    &&\tau_A\delta_{A\pm}
    \end{pmatrix}
    \underbrace{\begin{pmatrix}
       \sqrt{3}\\ -1 \\-\sqrt{2}
   \end{pmatrix}}_\Bb
\end{align}
where the left side is read off the three-nucleon Lagrangian in Eq.~\eqref{eq:L_3body} without the factor of $\sqrt{4\pi/M_N}$.
The projector in Eq.~\eqref{eq:3Ndoublet-projector} and its Hermitian conjugate can be rescaled to recover the non-symmetric matrices  used in Refs.~\cite{Ando:2010wq,Vanasse:2014kxa,Konig:2014ufa,Konig:2015aka} that correspond to $\Xb$ in this work. To maintain the normalization of $\mathbf{P}^\dagger\mathbf{P}$, the rescaling can be constructed as
\begin{align}
    \mathbf{P} \to \mathbf{S} \mathbf{P}, \qquad \mathbf{P}^\dagger \to  \mathbf{P}^\dagger\mathbf{S}^{-1}.
\end{align}
As an example, one could choose $\mathbf{S} = \textrm{diag}(1,\sqrt{1/3},\sqrt{2/3})$, in which case $\Xb$ becomes
\begin{align}
    \Xb \to \mathbf{S}\Xb\mathbf{S}^{-1} = -\frac{1}{8}\begin{pmatrix}
        -1 & 3&3\\
        1 & 1& -1\\
        2 & -2 & 0
    \end{pmatrix},
\end{align}
which is used in Ref.~\cite{Konig:2015aka}.

\subsection{Type-(a) and Type-(b) Contractions}

For the one-nucleon isoscalar magnetic current, the \ac{CCS} matrices are
\begin{align}
\Mb^{s}_{a,1} &= \Mb^{s}_{a,2}=
    \text{diag}\!\left(\frac{1}{3}, 0, 0\right)\kappa_0,
    \qquad
\Mb^{s}_{a,3} =
    \text{diag}\!\left(-\frac{1}{6}, \frac{1}{2}, \frac{1}{2}\right)\kappa_0,\\
\Mb^{s}_{b,1} &= (\Mb^{s}_{b,3})^\intercal=  4\Xb\Mb^{s}_{a,3}\kappa_0 ,
    \qquad
\Mb^{s}_{b,2} = -2\cdot\frac{1}{8}
    \begin{pmatrix}
    -5/3 & 1/\sqrt{3} & \sqrt{2/3} \\
    1/\sqrt{3} & -1 & \sqrt{2} \\
    \sqrt{2/3} & \sqrt{2} & 0
\end{pmatrix}\kappa_0,
\end{align}
where all type-($b$) matrices receive a factor of 2 because both permutations in $\hat{P} = P_{12}P_{23} + P_{13}P_{23}$ contribute to the first term in Eq.~\eqref{eq:psi-O-psi} through the type-($b$) contraction in Fig.~\ref{fig:He3formfactor}. 

For the one-nucleon isovector magnetic current, the \ac{CCS} matrices are
\begin{align}
\Mb^{v}_{a,1}&=  \Mb^{v}_{a,2}=
    \begin{pmatrix}
        0& 1/2\sqrt{3}&0\\
       1/2\sqrt{3}& 0&0\\
        0&0& 0
    \end{pmatrix}\kappa_1,
    \qquad
\Mb^{v}_{a,3} =
    \text{diag}\!\left(-\frac{1}{6}, \frac{1}{2}, -\frac{1}{2}\right)\kappa_1,\\
\Mb^{v}_{b,1} &= (\Mb^{v}_{b,3})^\intercal=  4\Xb\Mb^{v}_{a,3} \kappa_1,
    \qquad
\Mb^{v}_{b,2} = -2\cdot\frac{1}{8}
    \begin{pmatrix}
        5/3& -\sqrt{1/3}&\sqrt{2/3}\\
        -\sqrt{1/3}& 1&\sqrt{2}\\
        \sqrt{2/3}&\sqrt{2}& 0
    \end{pmatrix}\kappa_1.
\end{align}

For the one-nucleon $E1$ current, the \ac{CCS} matrices are
\begin{align}
\Mb^{C}_{a,1} &= \Mb^{C}_{a,2}=
    \text{diag}\!\left(\frac{1}{4}, \frac{1}{4}, \frac{1}{2}\right),
    \qquad
\Mb^{C}_{a,3} =
    \text{diag}\!\left(\frac{1}{2}, \frac{1}{2},0\right),\\
\Mb^{C}_{b,1} &= (\Mb^{C}_{b,3})^\intercal=  4\Xb\Mb^{C}_{a,3} ,
    \qquad
\Mb^{C}_{b,2} = -2\cdot\frac{1}{8}
    \begin{pmatrix}
    0 & 0& \sqrt{6} \\
    0 & 0 & -\sqrt{2} \\
    \sqrt{6} &  -\sqrt{2} & 0
\end{pmatrix}.
\end{align}

For  the one-nucleon scalar current, the \ac{CCS} matrices are
\begin{align}
    \Mb^{\#}_{a,i} = \frac{\boldsymbol{\mathcal{I}}}{2}, \quad \Mb^{\#}_{b,i} = 2\Xb,
\end{align}
where $\mathcal{I}$ is the $3\times 3$ identity matrix in the three-channel formalism.

\subsection{Type-(c) Contraction}
For the one-nucleon magnetic isoscalar current, the \ac{CCS} matrices are
\begin{align}
\Mb^{s}_{c,1} &= 
    \text{diag}\!\left(-\frac{1}{12}, \frac{1}{4}, 0\right)\kappa_0,
\end{align}
and
\begin{align}
    \Mb^{s}_{c,2} =
    \begin{pmatrix}
    1/6 & 1/4\sqrt{3} & 0 \\
    1/4\sqrt{3} & 0 & 0 \\
    0 & 0 & 0
\end{pmatrix}\kappa_0, \qquad
\Mb^{s}_{c,3} =
    \begin{pmatrix}
    1/6 & -1/4\sqrt{3} & 0 \\
    -1/4\sqrt{3} & 0 & 0 \\
    0 & 0 & 0
\end{pmatrix}\kappa_0.
\end{align}
The \ac{CCS} matrices for the one-nucleon isovector magnetic current are related to those for the isoscalar one by
\begin{align}
    \Mb^{v}_{c,i} &= \Mb^{s}_{c,i}\frac{\kappa_1}{\kappa_0}, \text{ for }i = 1,2; \qquad \Mb^{v}_{c,3} = -\Mb^{s}_{c,3}\frac{\kappa_1}{\kappa_0}.
\end{align}
For the one-nucleon $E1$ current, the \ac{CCS} matrices are
\begin{align}
   \Mb^{C}_{c,1} = \Mb^{C}_{c,2} = \text{diag}\!\left(\frac{1}{4}, \frac{1}{4}, 0\right), \qquad
    \Mb^{C}_{c,3} = 0.
\end{align}
For the one-nucleon scalar current, the \ac{CCS} matrices are
\begin{align}
   \Mb^{\#}_{c,1} = \Mb^{\#}_{c,2} = \Mb^{\#}_{c,3} = \frac{\mathcal{I}}{4}.
\end{align}

\subsection{Type-(d), (e), and (f) Contractions}
$\Mb^\xi_{\alpha,i}$ for $\alpha = d, e, f$ can be constructed using the matrices above along with the diagonal matrix,
\begin{align}
    \boldsymbol{\delta}_{33} = \text{diag}\!\left(0, 0, 1\right),
\end{align}
which is used to project out the \ac{pp} channel for $\He$. The construction works the same for any of the currents considered in this work. A superscript $O =  s, v, C, \#$ will thus be used below. For the type-$(d)$ permutation,
\begin{align}
    \Mb^\xi_{d,1}&= (\mathcal{I} - \boldsymbol{\delta}_{33})2\Mb^\xi_{a,3}\Xb\boldsymbol{\delta}_{33}, 
    \\
    \Mb^\xi_{d,2}&= (\mathcal{I} - \boldsymbol{\delta}_{33})\frac{1}{2}\Mb^\xi_{b,2}\boldsymbol{\delta}_{33},
    \\
    \Mb^\xi_{d,3}&= (\mathcal{I} - \boldsymbol{\delta}_{33})2\Xb\Mb^\xi_{a,3}\boldsymbol{\delta}_{33}.
\end{align}
For the type-$(e)$ permutation, the \ac{CCS} matrices are
\begin{align}
    \Mb^\xi_{e,1}&= (\mathcal{I} - \boldsymbol{\delta}_{33})2\Mb^\xi_{a,3}\Xb(\mathcal{I} - \boldsymbol{\delta}_{33}), 
    \\
    \Mb^\xi_{e,2}&= (\mathcal{I} - \boldsymbol{\delta}_{33})2\Xb\Mb^\xi_{a,3}(\mathcal{I} - \boldsymbol{\delta}_{33}),
    \\
     \Mb^\xi_{e,3}&= (\mathcal{I} - \boldsymbol{\delta}_{33})\frac{1}{2}\Mb^\xi_{b,2}(\mathcal{I} - \boldsymbol{\delta}_{33}).
\end{align}
For the type-$(f)$ permutation, the \ac{CCS} matrices are
\begin{align}
    \Mb^\xi_{f,i}&= \Mb^\xi_{c,i}, \text{ for } i = 1,2,3.
\end{align}

\section{Relation between three-body vertex and wavefunction}
\label{app:vertex2wave}
This appendix briefly reviews the Faddeev formalism for three-body systems~\cite{Faddeev:1960su} and relates the three-body vertex to the wavefunction. This can be used to obtain the $\He$ wavefunction from the vertex function that has been studied in, e.g., Refs.~\cite{Ando:2010wq,Konig:2015aka}.

\subsection{With separable two-body interaction}
A separable two-body potential embedded in a three-body system can be written as
\begin{align}
    \hat{V}_s &= \left(|\phi\rangle v_2 \langle\phi|\right) \otimes \hat{\mathcal{I}}_{p_2}, \qquad \hat{\mathcal{I}}_{p_2} = \left(\sum_{\vect{p}_2}|\vect{p}_2\rangle\langle\vect{p}_2|\right) 
\end{align}
where $\vect{p}_1$ and $\vect{p}_2$ are Jacobi momenta, with $\vect{p}_1$ being the relative momentum between two particles interacting through $|\phi\rangle v_2 \langle\phi|$, and $\vect{p}_2$ being the relative momentum between the interacting two-body subsystem and the spectator. In principle one needs to permute $\hat{V}_s$ as the interaction can apply to any of the three pairs. However, in the Faddeev formalism this is simplified and one only needs to consider one pair.
$|\phi\rangle$ is the regulator state defined as
\begin{align}
\label{eq:regulator_phi}
    |\phi\rangle = \sum_{\vect{p}_1} \phi(p_1)|\vect{p}_1\rangle,
\end{align}
where $\phi(p_1)$ is a generic two-body regulator in momentum space.
The separable two-body $T$-matrix $\hat{t}_2$ embeded in the three-body space can be written as
\begin{align}
    \label{eq:t2}
    \hat{t}_2 &= \hat{V}_s(1-\hat{G}_0\hat{V}_s)^{-1}\nonumber\\
    &=\hat{\tau}_2\otimes\hat{\mathcal{I}}_{p_2}
\end{align}
where $\hat{G}_0$ is the free three-body Green's function and $\hat{\tau}_2$ is given by
\begin{align}
    \hat{\tau}_2 = \frac{|\phi\rangle v_2 \langle \phi |}{1 - \langle\phi|\hat{G}_0\hat{V}_s|\phi\rangle}.
\end{align}

The wavefunction for three identical particles  can be decomposed using
\begin{align}
    |\Psi\rangle = \frac{1 + \hat{P}}{3}|\psi\rangle, 
\end{align}
where $\hat{P} = \hat{P}_{12}\hat{P}_{23} + \hat{P}_{31}\hat{P}_{23}$ is the permutation operator and $(1 + \hat{P})$ generates the cyclic permutations of the three particles. $|\psi\rangle$ is given by the solution to the Faddeev equation,
\begin{align}
    \label{eq:Faddeev}
    |\psi\rangle = \hat{G}_0\hat{t}_2\hat{P}|\psi\rangle
\end{align}
The three-body vertex function can be defined as 
\begin{align}
\label{eq:vertex_state}
    |\mathcal{G}\rangle &\equiv \langle\phi|\hat{P}|\psi\rangle,
\end{align}
which is still a vector in the $|\vect{p}_2\rangle$ space.
Comparing Eqs.~\eqref{eq:Faddeev} and~\eqref{eq:vertex_state} shows that the solution of Eq.~\eqref{eq:Faddeev} can be written as
\begin{align}
    \label{eq:3Bvertex_to_wavefunction}
    |\psi\rangle = \hat{G}_0\left( \hat{
    \tau
    }_2 |\mathcal{G}\rangle\otimes|\phi\rangle \right).
\end{align}
Plugging \eqref{eq:3Bvertex_to_wavefunction} into Eq.~\eqref{eq:vertex_state} gives the integral equation for $|\mathcal{G}\rangle$,
\begin{align}
    |\mathcal{G}\rangle = 
    \hat{K}|\mathcal{G}\rangle,
\end{align}
where the kernel
\begin{align}
    \hat{K} = \langle \phi|\hat{P}\hat{G}_0|\phi \rangle \,\hat{\tau}_2
\end{align}
is an operator in the $|\vect{p}_2\rangle$ space and has the same form as Eq.~\eqref{eq:kernel_strong} up to spin-isospin structures.

\subsection{Including non-separable two-body interaction}
In the presence of both separable and non-separable two-body interactions, such as the strong plus the Coulomb interaction, the two-body $T$-matrix can be written as the sum of a separable part, denoted $\hat{t}_{sc}$, and a non-separable piece, denoted $\hat{t}_{c}$. Denote the separable two-body interaction as $\hat{V}_s$ and the non-separable one, $\hat{V}_c$. The two-body $T$-matrix is given by
\begin{align}
    \hat{t}_2 &= \left(\hat{V}_s + \hat{V}_c\right) \left(1 - \hat{G}_0\left(\hat{V}_s + \hat{V}_c\right)\right)^{-1} \\
    & \equiv \hat{t}_{sc}+ \hat{t}_{c},
\end{align}
where
\begin{align}
\label{eq:tc}
    \hat{t}_{c} &= \hat{V}_c \left(1 - \hat{G}_0\hat{V}_c\right)^{-1},\\
\label{eq:tsc}
    \hat{t}_{sc} &= (1 + \hat{t}_{c}\hat{G}_0)\hat{\tau}_{sc}(1 + \hat{G}_0\hat{t}_{c}),
\end{align}
and 
\begin{align}
    \hat{\tau}_{sc} &= \frac{|\phi\rangle v_2\langle\phi|}{1 - \left\langle\phi\left|(1+\hat{G}_0\hat{t}_c)\hat{G}_0\hat{V}_s\right|\phi\right\rangle}.
\end{align}

The three-body Faddeev equation now reads
\begin{align}
    |\Psi\rangle &= \frac{1+\hat{P}}{3} |\psi\rangle\\
    |\psi\rangle &= \underbrace{\hat{G}_0\hat{t}_{sc}\hat{P}|\psi\rangle}_{\text{\normalsize$|\psi_{sc}\rangle$}} + \underbrace{\hat{G}_0\hat{t}_{c}\hat{P}|\psi\rangle}_{\text{\normalsize $|\psi_{c}\rangle$}}\\
\end{align}
where $|\psi_{sc}\rangle$ and $|\psi_{c}\rangle$ satisfy
\begin{align}
\label{eq:Faddeev_sc_c}
    \begin{pmatrix}
        |\psi_{sc}\rangle\\
        |\psi_{c}\rangle 
    \end{pmatrix}
    = \hat{G}_0\begin{pmatrix}
        \hat{t}_{sc} &  \hat{t}_{sc} \\
        \hat{t}_{c} &  \hat{t}_{c}
    \end{pmatrix}\hat{P}\begin{pmatrix}
        |\psi_{sc}\rangle\\
        |\psi_{c}\rangle
    \end{pmatrix}.
\end{align}
$|\psi_{sc}\rangle$ is thus given by
\begin{align}
    \label{eq:Faddeev_sc}
    |\psi_{sc}\rangle = \hat{G}_0\hat{t}_{sc}\hat{P}\left(1 + \hat{T}_c\right)|\psi_{sc}\rangle,
\end{align}
where 
\begin{align}
    \hat{T}_c = \left(1 - \hat{G}_0\hat{t}_{c}\hat{P}\right)^{-1}\hat{G}_0\hat{t}_{c}\hat{P}.
\end{align}
Note that once $|\psi_{sc}\rangle$ is solved for, $|\psi_c\rangle$ can be obtained using
\begin{align}
\label{eq:psi_c}
     |\psi_{c}\rangle  &= \hat{T}_c|\psi_{sc}\rangle.
\end{align}
The three-body vertex function can now be defined as
\begin{align}
\label{eq:3Bvertex_sc}
    |\mathcal{G}\rangle &\equiv \langle\phi|(1+\hat{G}_0\hat{t}_c)\hat{P}(1+\hat{T}_c)|\psi_{sc}\rangle.
\end{align}
Using Eqs.~\eqref{eq:tsc} and~\eqref{eq:Faddeev_sc}, $|\psi_{sc}\rangle$ can be written as
\begin{align}
    \label{eq:3Bvertex_to_wavefunction_sc}
    |\psi_{sc}\rangle = \hat{G}_0\left(1+\hat{t}_c\hat{G}_0\right)\left( \hat{\tau}_{sc} |\mathcal{G}\rangle\otimes|\phi\rangle \right).
\end{align}
Plugging Eq.~\eqref{eq:3Bvertex_to_wavefunction_sc} into Eq.~\eqref{eq:3Bvertex_sc} gives the integral equation for $|\mathcal{G}\rangle$,
\begin{align}
    |\mathcal{G}\rangle &= \hat{K}|\mathcal{G}\rangle,\qquad 
    \hat{K} = \left\langle \phi\left|(1+\hat{G}_0\hat{t}_c)\hat{P}(1+\hat{T}_c)\hat{G}_0(1+\hat{t}_c\hat{G}_0)\right|\phi \right\rangle \,\hat{\tau}.
    \label{eq:K_sc}
\end{align}

\subsection{Including separable three-body interaction}
Consider a momentum-independent separable three-body interaction $\hat{V}_3$, 
\begin{align}
     \hat{V}_3 = |\xi\rangle v_3 \langle\xi|,
\end{align}
where $|\xi\rangle$ is given by
\begin{align}
    |\xi\rangle = |\phi\rangle\otimes|\phi_3\rangle = \sum_{\vect{p}_1,\vect{p}_2} \phi(p_1)\phi_3(p_2)|\vect{p}_1 \vect{p}_2\rangle.
\end{align}
In the above equation $\phi(p_1)$ and $\phi_3(p_2)$ are regulators for each Jacobi momentum, with the former being the same as the two-body one in Eq.~\eqref{eq:regulator_phi}.
The Faddeev equation with $\hat{V}_3 $ can be written as~\cite{Stadler:1991ed}
\begin{align}
    \label{eq:Faddeev_w3BF_1}
    |\psi\rangle &= \hat{G}_0\left(\hat{t}_{sc} + \hat{t}_{c}\right)\hat{P}|\psi\rangle  + \hat{G}_0\left[1 + \left(\hat{t}_{sc} + \hat{t}_{c}\right) \hat{G}_0\right]\hat{V}_3(\cdots)|\psi\rangle
\end{align}
The ellipsis above contains a sequence of operators between $\hat{V}_3$ and $|\psi\rangle$. However, since $\hat{V}_3$ is separable, everything on its right side can be absorbed into a number. This number can be fixed by the normalization of the three-body wavefunction and does not need to be written out explicitly here.  Up to a rescaling of the inhomogeneous term (i.e., the last term in Eq.~\eqref{eq:Faddeev_w3BF_1}), Eq.~\eqref{eq:Faddeev_w3BF_1} can be written as
\begin{align}
    \label{eq:Faddeev_w3BF_2}
    |\psi\rangle & = \hat{G}_0\left(\hat{t}_{sc} + \hat{t}_{c}\right)\hat{P}|\psi\rangle + \hat{G}_0\left(\hat{t}_{sc} + \hat{t}_{c}\right)\hat{G}_0 {v_2}|\xi\rangle\nonumber\\
    &=\underbrace{\hat{G}_0\hat{t}_{sc}\hat{P}|\psi\rangle + \hat{G}_0(1+\hat{t}_c\hat{G}_0)\hat{\tau}_{sc}\Big(|\phi\rangle\otimes|\phi_3\rangle \Big)}_{\text{\normalsize$|\psi_{sc}\rangle$}}
     + \underbrace{\hat{G}_0\hat{t}_{c}\hat{P}|\psi\rangle \vphantom{ \hat{G}_0(1+\hat{t}_c\hat{G}_0)\hat{\tau}_{sc}\Big(|\phi\rangle\otimes|\phi_3\rangle \Big) }   }_{\text{\normalsize$|\psi_{c}\rangle$}}
\end{align}
where the second line is obtained using Eqs.~\eqref{eq:tc} and~\eqref{eq:tsc}. $|\psi_c\rangle$ and $|\psi_{sc}\rangle$ are defined similarly to Eq.~\eqref{eq:Faddeev_sc_c}, except that the latter now receives an inhomogeneous term and becomes
\begin{align}
\label{eq:Faddeev_w3BF_3}
        \begin{pmatrix}
        |\psi_{sc}\rangle\\
        |\psi_{c}\rangle 
    \end{pmatrix}
    = \hat{G}_0\begin{pmatrix}
        \hat{t}_{sc} &  \hat{t}_{sc} \\
        \hat{t}_{c} &  \hat{t}_{c}
    \end{pmatrix}\hat{P}\begin{pmatrix}
        |\psi_{sc}\rangle\\
        |\psi_{c}\rangle
    \end{pmatrix} 
    +\begin{pmatrix}
        \hat{G}_0(1+\hat{t}_c\hat{G}_0)\hat{\tau}_{sc}\Big(|\phi\rangle\otimes|\phi_3\rangle \Big)\\
        0
    \end{pmatrix}.
\end{align}
$|\psi_{sc}\rangle$ now satisfies
\begin{align}
    \label{eq:Faddeev_sc_w3BF}
    |\psi_{sc}\rangle = \hat{G}_0\hat{t}_{sc}\hat{P}\left(1 + \hat{T}_c\right)|\psi_{sc}\rangle +  \hat{G}_0(1+\hat{t}_c\hat{G}_0)\hat{\tau}_{sc}\Big(|\phi\rangle\otimes|\phi_3\rangle \Big)
\end{align}
One might attempt to follow Eq.~\eqref{eq:3Bvertex_sc} and define the three-body vertex function as
\begin{align}
        |\widetilde{\mathcal{G}}\rangle &\equiv \langle\phi|(1+\hat{t}_c)\hat{P}(1+\hat{T}_c)|\psi_{sc}\rangle\\
         &= \hat{K}|\widetilde{\mathcal{G}}\rangle + \hat{K}|\phi_3\rangle,
         \label{eq:G_sc_tilde}
\end{align}
where $\hat{K}$ is the same as that defined in Eq.~\eqref{eq:K_sc} and does not involve a three-body force. However, the structure of Eq~\eqref{eq:G_sc_tilde} suggests a more convenient definition of the vertex function
\begin{align}
    |{\mathcal{G}}\rangle &= \hat{K}|{\mathcal{G}}_{sc}\rangle +|\phi_3\rangle,
\end{align}
which, up to spin-isospin structures, is equivalent to Eq.~\eqref{eq:vertex_function} by choosing the regulator $\phi_3(p_2) = \theta(\Lambda-p_2)$ as a step function. Eqs.~\eqref{eq:Faddeev_w3BF_3} and~\eqref{eq:Faddeev_sc_w3BF} are solved by
\begin{align}
    |\psi_{sc}\rangle &= \hat{G}_0\left(1+\hat{t}_c\hat{G}_0\right)\left( \hat{\tau}_{sc} |\mathcal{G}\rangle\otimes|\phi\rangle \right)\\
         |\psi_{c}\rangle  &= \left(1-\hat{G}_0\hat{t}_c\hat{P}\right)^{-1}\hat{G}_0\hat{t}_c\hat{P}|\psi_{sc}\rangle.
    \label{eq:3Bvertex_to_wavefunction_final}
\end{align}
Here it is assumed that the non-separable two-body interaction apply to all pairs of the three particles. If only two can interact this way (e.g., two electrically charged and another neutral particles), then the expression in Eq.~\eqref{eq:3Bvertex_to_wavefunction_final} can be simiplified by only keeping only one permutation in $\hat{P}$. For example,  the expression $(1-\hat{G}_0\hat{t}_c\hat{P})^{-1}\hat{G}_0\hat{t}_c\hat{P}$ can be reduced to $\hat{G}_0\hat{t}_c\hat{P}$. For the $\He$ case, one also needs to include the spin-isospin structures of three-nucleon systems in the three-channel formalism. This results in  Eqs.~\eqref{eq:psi_sc_he3_tilde} and~\eqref{eq:psi_c_he3_tilde}, or equivalently their diagrammatic representations in Figs.~\ref{fig:psi_sc} and~\ref{fig:psi_c}.

\section{Interpolation}
\label{app:interpolation}
Assuming a function $f(q_1,q_2)$ only depends on the magnitude of three-dimensional vectors $\vect{q}_1$ and $\vect{q}_2$, its value at $f({p}_1, {p}_2)$ can be obtained by
\begin{align}
\label{eq:interpolating}
   f(p_1, p_2) = \int_0^{\infty}\frac{q_1^2dq_1}{2\pi^2}\int_0^{\infty}\frac{q_2^2dq_2}{2\pi^2}S(p_1, p_2; q_1,q_2)f({q_1},{q_2}),
\end{align}
where $S(p_1, p_2; q_1,q_2)$ is an interpolating function,
\begin{align}
\label{eq:S_cubic}
    S(p_1, p_2; q_1,q_2) = &C(p_1,q_1)C(p_2,q_2),\nonumber\\
    \sim &\delta(q_1-p_1)\delta(q_2 - p_2),
\end{align}
where $C(p_1,q_1)$ smears out $\delta(p_1,q_1)$. In this work $C(p_1,q_1)$ is chosen to be a one-dimensional cubic spline. In addition, derivatives of the delta function is replaced with those of the cubic spline, 
\begin{align}
    \frac{d}{dp}\delta(p'-p) &\sim \frac{d}{dp}C(p',p),
\end{align}
where the first derivative on the right-hand side can be computed analytically. The second derivative of the delta function is treated numerically as
\begin{align}
    \frac{d^2}{d^2p}\delta(p'-p) &\sim \int_{0}^\Lambda dq \frac{d}{dp}C(p,q)\frac{d}{dq}C(q,p'),
\end{align}
to avoid non-smoothness in the second derivative of the cubic spline.

\begin{acronym}[ChEFT]\itemsep1em  
\acro{CM}{center-of-mass}
\acro{ChEFT}{Chiral effective field theory}
\acro{CCS}{cluster configuration space}
\acro{pp}[$pp$]{proton-proton}
\acro{BSM}{beyond-the-standard-model}
\acro{DM}{dark matter}
\acro{EFT}{effective field theory}
\acrodefplural{EFT}{effective field theories}
\acro{ERE}{effective range expansion}
\acro{E1}{electric dipole}
\acro{E2}{electric quadruple}
\acro{FY}{Faddeev-Yakubovsky}
\acro{STM}{}
\acro{LO}{leading order}
\acro{LEC}{low-energy coefficient}
\acrodefplural{LEC}{low-energy coefficients}
\acro{MS}{minimal subtraction}
\acro{M1}{magnetic dipole}
\acro{NDA}{naive dimensional analysis}
\acro{nd}[$nd$]{neutron-deuteron}
\acro{a-nd}[$a_{nd}$]{neutron-deuteron scattering length}
\acro{np}[$np$]{neutron-proton}
\acro{NLO}{next-to-leading order}
\acro{NNLO}{next-to-next-to-leading order}
\acro{N3LO}{next-to-next-to-next-to-leading order}
\acro{Nc}[$N_c$]{number of QCD colors}
\acro{eftnopi}[\eftnopi]{pionless effective field theory}
\acro{PV}{parity-violating}
\acro{PDS}{power divergence subtraction}
\acro{QCD}{quantum chromodynamics}
\acro{RG}{renormalization group}
\acro{SREFT}{short-range effective field theory}
\acro{SD}{spin-dependent}
\acro{SI}{spin-independent}
\acro{WIMP}{weakly-interacting-massive particle}
\acrodefplural{WIMP}{weakly-interacting-massive particles}
\acro{triton}[${}^3\mathrm{H}$]{triton}
\acro{He}[${}^3\mathrm{He}$]{Helium-3}

\end{acronym}

\bibliography{ref}

@article{Vanasse:2013sda,
    author = "Vanasse, Jared",
    title = "{Fully Perturbative Calculation of $nd$ Scattering to Next-to-next-to-leading-order}",
    eprint = "1305.0283",
    archivePrefix = "arXiv",
    primaryClass = "nucl-th",
    doi = "10.1103/PhysRevC.88.044001",
    journal = "Phys. Rev. C",
    volume = "88",
    number = "4",
    pages = "044001",
    year = "2013"
}

@article{Vanasse:2015fph,
    author = "Vanasse, Jared",
    title = "{Triton charge radius to next-to-next-to-leading order in pionless effective field theory}",
    eprint = "1512.03805",
    archivePrefix = "arXiv",
    primaryClass = "nucl-th",
    doi = "10.1103/PhysRevC.95.024002",
    journal = "Phys. Rev. C",
    volume = "95",
    number = "2",
    pages = "024002",
    year = "2017"
}

@article{Griesshammer:2004pe,
    author = "Griesshammer, Harald W.",
    title = "{Improved convergence in the three-nucleon system at very low energies}",
    eprint = "nucl-th/0404073",
    archivePrefix = "arXiv",
    reportNumber = "TUM-T39-04-05",
    doi = "10.1016/j.nuclphysa.2004.08.012",
    journal = "Nucl. Phys. A",
    volume = "744",
    pages = "192--226",
    year = "2004"
}

@article{Kaplan:1998tg,
    author = "Kaplan, David B. and Savage, Martin J. and Wise, Mark B.",
    title = "{A New expansion for nucleon-nucleon interactions}",
    eprint = "nucl-th/9801034",
    archivePrefix = "arXiv",
    reportNumber = "DOE-ER-40561-352, INT-97-00-189, NT-UW-98-05, CALT-68-2155",
    doi = "10.1016/S0370-2693(98)00210-X",
    journal = "Phys. Lett. B",
    volume = "424",
    pages = "390--396",
    year = "1998"
}

@article{Chen:1999tn,
    author = "Chen, Jiunn-Wei and Rupak, Gautam and Savage, Martin J.",
    title = "{Nucleon-nucleon effective field theory without pions}",
    eprint = "nucl-th/9902056",
    archivePrefix = "arXiv",
    reportNumber = "NT-UW-99-14, JLAB-THY-99-50",
    doi = "10.1016/S0375-9474(99)00298-5",
    journal = "Nucl. Phys. A",
    volume = "653",
    pages = "386--412",
    year = "1999"
}

@Article{Schmidt1937,
    author="Schmidt, Th.",
    title="{\"U}ber die magnetischen Momente der Atomkerne",
    journal="Zeitschrift f{\"u}r Physik",
    year="1937",
    volume="106",
    number="5",
    pages="358--361",
    issn="0044-3328",
    doi="10.1007/BF01338744",
    url="http://dx.doi.org/10.1007/BF01338744"
}

@article{Vanasse:2017kgh,
    author = "Vanasse, Jared",
    title = "{Charge and Magnetic Properties of Three-Nucleon Systems in Pionless Effective Field Theory}",
    eprint = "1706.02665",
    archivePrefix = "arXiv",
    primaryClass = "nucl-th",
    doi = "10.1103/PhysRevC.98.034003",
    journal = "Phys. Rev. C",
    volume = "98",
    number = "3",
    pages = "034003",
    year = "2018"
}

@article{Bedaque:1999ve,
    author = "Bedaque, Paulo F. and Hammer, H.-W. and van Kolck, U.",
    title = "{Effective theory of the triton}",
    eprint = "nucl-th/9906032",
    archivePrefix = "arXiv",
    reportNumber = "DOE-ER-40561-57, TRI-PP-99-24, KRL-MAP-248, NT-UW-99-30",
    doi = "10.1016/S0375-9474(00)00205-0",
    journal = "Nucl. Phys. A",
    volume = "676",
    pages = "357--370",
    year = "2000"
}

@article{Bedaque:2002yg,
    author = "Bedaque, Paulo F. and Rupak, Gautam and Griesshammer, Harald W. and Hammer, Hans-Werner",
    title = "{Low-energy expansion in the three-body system to all orders and the triton channel}",
    eprint = "nucl-th/0207034",
    archivePrefix = "arXiv",
    reportNumber = "LBL-50137, TUM-T39-02-12",
    doi = "10.1016/S0375-9474(02)01402-1",
    journal = "Nucl. Phys. A",
    volume = "714",
    pages = "589--610",
    year = "2003"
}

@article{Kirscher:2017fqc,
    author = "Kirscher, Johannes and Pazy, Ehoud and Drachman, Jonathan and Barnea, Nir",
    title = "{Electromagnetic characteristics of $A \leq 3$ physical and lattice nuclei}",
    eprint = "1702.07268",
    archivePrefix = "arXiv",
    primaryClass = "nucl-th",
    doi = "10.1103/PhysRevC.96.024001",
    journal = "Phys. Rev. C",
    volume = "96",
    number = "2",
    pages = "024001",
    year = "2017"
}

@article{Vanasse:2016umz,
    author = "Vanasse, Jared and Phillips, Daniel R.",
    title = "{Three-nucleon bound states and the Wigner-SU(4) limit}",
    eprint = "1607.08585",
    archivePrefix = "arXiv",
    primaryClass = "nucl-th",
    doi = "10.1007/s00601-016-1173-2",
    journal = "Few Body Syst.",
    volume = "58",
    number = "2",
    pages = "26",
    year = "2017"
}

@article{Bedaque:1998mb,
    author = "Bedaque, Paulo F. and Hammer, H.-W. and van Kolck, U.",
    title = "{Effective theory for neutron deuteron scattering: Energy dependence}",
    eprint = "nucl-th/9802057",
    archivePrefix = "arXiv",
    reportNumber = "DOE-ER-40561-356, INT-98-00-4, TRI-PP-98-2, KRL-MAP-219",
    doi = "10.1103/PhysRevC.58.R641",
    journal = "Phys. Rev. C",
    volume = "58",
    pages = "R641--R644",
    year = "1998"
}

@article{Gabbiani:1999yv,
    author = "Gabbiani, Fabrizio and Bedaque, Paulo F. and Griesshammer, Harald W.",
    title = "{Higher partial waves in an effective field theory approach to nd scattering}",
    eprint = "nucl-th/9911034",
    archivePrefix = "arXiv",
    reportNumber = "DOE-ER-40561-74, DUKE-TH-99-198, NT-UW-99-61, TUM-T39-99-23",
    doi = "10.1016/S0375-9474(00)00181-0",
    journal = "Nucl. Phys. A",
    volume = "675",
    pages = "601--620",
    year = "2000"
}

@article{Rupak:2001ci,
    author = "Rupak, Gautam and Kong, Xin-wei",
    title = "{Quartet S wave p d scattering in EFT}",
    eprint = "nucl-th/0108059",
    archivePrefix = "arXiv",
    reportNumber = "TRI-PP-01-13",
    doi = "10.1016/S0375-9474(03)00638-9",
    journal = "Nucl. Phys. A",
    volume = "717",
    pages = "73--90",
    year = "2003"
}

@article{Sadeghi:2006fc,
    author = "Sadeghi, H. and Bayegan, S. and Griesshammer, Harald W.",
    title = "{Effective field theory calculation of thermal energies and radiative capture cross-section}",
    eprint = "nucl-th/0610029",
    archivePrefix = "arXiv",
    doi = "10.1016/j.physletb.2006.10.023",
    journal = "Phys. Lett. B",
    volume = "643",
    pages = "263",
    year = "2006"
}

@article{Arani:2014qsa,
    author = "Arani, M. Moeini and Nematollahi, H. and Mahboubi, N. and Bayegan, S.",
    title = "{New insight into the $nd\!\!\to\!\!{}^3\mathrm{H}\gamma$ process at thermal energy with pionless effective field theory}",
    eprint = "1406.6530",
    archivePrefix = "arXiv",
    primaryClass = "nucl-th",
    doi = "10.1103/PhysRevC.89.064005",
    journal = "Phys. Rev. C",
    volume = "89",
    number = "6",
    pages = "064005",
    year = "2014"
}

@article{Rupak:1999rk,
    author = "Rupak, Gautam",
    title = "{Precision calculation of $np\to d\gamma$ cross-section for big bang nucleosynthesis}",
    eprint = "nucl-th/9911018",
    archivePrefix = "arXiv",
    reportNumber = "NT-UW-99-54",
    doi = "10.1016/S0375-9474(00)00323-7",
    journal = "Nucl. Phys. A",
    volume = "678",
    pages = "405--423",
    year = "2000"
}

@article{Konig:2013cia,
    author = {K\"onig, Sebastian and Hammer, H.-W.},
    title = "{Precision calculation of the quartet-channel p-d scattering length}",
    eprint = "1312.2573",
    archivePrefix = "arXiv",
    primaryClass = "nucl-th",
    doi = "10.1103/PhysRevC.90.034005",
    journal = "Phys. Rev. C",
    volume = "90",
    number = "3",
    pages = "034005",
    year = "2014"
}

@article{Koenig:2011lmm,
    author = {K\"onig, Sebastian and Hammer, H.-W.},
    title = "{Low-energy p-d scattering and He-3 in pionless EFT}",
    eprint = "1101.5939",
    archivePrefix = "arXiv",
    primaryClass = "nucl-th",
    doi = "10.1103/PhysRevC.83.064001",
    journal = "Phys. Rev. C",
    volume = "83",
    pages = "064001",
    year = "2011"
}

@article{Vanasse:2014kxa,
    author = {Vanasse, Jared and Egolf, David A. and Kerin, John and K\"onig, Sebastian and Springer, Roxanne P.},
    title = "{${}^{3}\mathrm{He}$ and $pd$ Scattering to Next-to-Leading Order in Pionless Effective Field Theory}",
    eprint = "1402.5441",
    archivePrefix = "arXiv",
    primaryClass = "nucl-th",
    doi = "10.1103/PhysRevC.89.064003",
    journal = "Phys. Rev. C",
    volume = "89",
    number = "6",
    pages = "064003",
    year = "2014"
}

@article{Konig:2014ufa,
    author = {K\"onig, Sebastian and Grie\ss{}hammer, Harald W. and Hammer, H.-W.},
    title = "{The proton-deuteron system in pionless EFT revisited}",
    eprint = "1405.7961",
    archivePrefix = "arXiv",
    primaryClass = "nucl-th",
    doi = "10.1088/0954-3899/42/4/045101",
    journal = "J. Phys. G",
    volume = "42",
    pages = "045101",
    year = "2015"
}

@article{Konig:2016iny,
    author = {K\"onig, Sebastian},
    title = "{Second-order perturbation theory for ${}^{3}\mathrm{He}$ and pd scattering in pionless EFT}",
    eprint = "1609.03163",
    archivePrefix = "arXiv",
    primaryClass = "nucl-th",
    doi = "10.1088/1361-6471/aa60d6",
    journal = "J. Phys. G",
    volume = "44",
    number = "6",
    pages = "064007",
    year = "2017"
}

@article{Wigner:1936dx,
    author = "Wigner, E.",
    title = "{On the Consequences of the Symmetry of the Nuclear Hamiltonian on the Spectroscopy of Nuclei}",
    doi = "10.1103/PhysRev.51.106",
    journal = "Phys. Rev.",
    volume = "51",
    pages = "106--119",
    year = "1937"
}

@article{Lin:2022yaf,
    author = "Lin, Xincheng and Singh, Hersh and Springer, Roxanne P. and Vanasse, Jared",
    title = "{Cold neutron-deuteron capture and Wigner-SU(4) symmetry}",
    eprint = "2210.15650",
    archivePrefix = "arXiv",
    primaryClass = "nucl-th",
    reportNumber = "INT-PUB-22-029",
    doi = "10.1103/PhysRevC.108.044001",
    journal = "Phys. Rev. C",
    volume = "108",
    number = "4",
    pages = "044001",
    year = "2023"
}

@article{Chen:1999vd,
    author = "Chen, Jiunn-Wei and Rupak, Gautam and Savage, Martin J.",
    title = "{Suppressed amplitudes in n p ---{\ensuremath{>}} d gamma}",
    eprint = "nucl-th/9905002",
    archivePrefix = "arXiv",
    reportNumber = "NT-UW-99-20",
    doi = "10.1016/S0370-2693(99)01007-2",
    journal = "Phys. Lett. B",
    volume = "464",
    pages = "1--11",
    year = "1999"
}

@article{Chen:1999bg,
    author = "Chen, Jiunn-Wei and Savage, Martin J.",
    title = "{n p ---{\ensuremath{>}} d gamma for big bang nucleosynthesis}",
    eprint = "nucl-th/9907042",
    archivePrefix = "arXiv",
    reportNumber = "NT-UW-99-35",
    doi = "10.1103/PhysRevC.60.065205",
    journal = "Phys. Rev. C",
    volume = "60",
    pages = "065205",
    year = "1999"
}

@article{Hammer:2019poc,
    author = {Hammer, H. -W. and K\"onig, S. and van Kolck, U.},
    title = "{Nuclear effective field theory: status and perspectives}",
    eprint = "1906.12122",
    archivePrefix = "arXiv",
    primaryClass = "nucl-th",
    doi = "10.1103/RevModPhys.92.025004",
    journal = "Rev. Mod. Phys.",
    volume = "92",
    number = "2",
    pages = "025004",
    year = "2020"
}

@article{Kirscher:2010dgl,
    author = "Kirscher, Johannes and Griesshammer, Harald W. and Shukla, Deepshikha and Hofmann, Hartmut M.",
    title = "{Universal Correlations in Pion-less EFT with the Resonating Group Model: Three and Four Nucleons}",
    eprint = "0903.5538",
    archivePrefix = "arXiv",
    primaryClass = "nucl-th",
    doi = "10.1140/epja/i2010-10939-5",
    journal = "Eur. Phys. J. A",
    volume = "44",
    pages = "239--256",
    year = "2010"
}

@article{Konig:2019xxk,
    author = {K\"onig, Sebastian},
    title = "{Energies and radii of light nuclei around unitarity}",
    eprint = "1910.12627",
    archivePrefix = "arXiv",
    primaryClass = "nucl-th",
    doi = "10.1140/epja/s10050-020-00098-9",
    journal = "Eur. Phys. J. A",
    volume = "56",
    number = "4",
    pages = "113",
    year = "2020",
}

@article{Stadler:1991ed,
  title = {Faddeev equations with three-nucleon force in momentum space},
  author = {Stadler, A. and Gl\"ockle, W. and Sauer, P. U.},
  journal = {Phys. Rev. C},
  volume = {44},
  issue = {6},
  pages = {2319--2327},
  numpages = {0},
  year = {1991},
  month = {Dec},
  publisher = {American Physical Society},
  doi = {10.1103/PhysRevC.44.2319},
  url = {https://link.aps.org/doi/10.1103/PhysRevC.44.2319}
}

@article{Ando:2010wq,
    author = "Ando, Shung-ichi and Birse, Michael C.",
    title = "{Effective field theory of 3He}",
    eprint = "1003.4383",
    archivePrefix = "arXiv",
    primaryClass = "nucl-th",
    doi = "10.1088/0954-3899/37/10/105108",
    journal = "J. Phys. G",
    volume = "37",
    pages = "105108",
    year = "2010"
}

@article{Konig:2015aka,
    author = {K{\"o}nig, Sebastian and Grie{\ss}hammer, Harald W. and Hammer, H. W. and van Kolck, U.},
    title = "{Effective theory of $^3$H and $^3$He}",
    eprint = "1508.05085",
    archivePrefix = "arXiv",
    primaryClass = "nucl-th",
    doi = "10.1088/0954-3899/43/5/055106",
    journal = "J. Phys. G",
    volume = "43",
    number = "5",
    pages = "055106",
    year = "2016"
}

@article{ParticleDataGroup:2024cfk,
    author = "Navas, S. and others",
    collaboration = "Particle Data Group",
    title = "{Review of particle physics}",
    doi = "10.1103/PhysRevD.110.030001",
    journal = "Phys. Rev. D",
    volume = "110",
    number = "3",
    pages = "030001",
    year = "2024"
}

@article{Lee:2015jqa,
    author = "Lee, Gabriel and Arrington, John R. and Hill, Richard J.",
    title = "{Extraction of the proton radius from electron-proton scattering data}",
    eprint = "1505.01489",
    archivePrefix = "arXiv",
    primaryClass = "hep-ph",
    reportNumber = "EFI-PREPRINT-14-35",
    doi = "10.1103/PhysRevD.92.013013",
    journal = "Phys. Rev. D",
    volume = "92",
    number = "1",
    pages = "013013",
    year = "2015"
}

@article{Epstein:2014zua,
    author = "Epstein, Zachary and Paz, Gil and Roy, Joydeep",
    title = "{Model independent extraction of the proton magnetic radius from electron scattering}",
    eprint = "1407.5683",
    archivePrefix = "arXiv",
    primaryClass = "hep-ph",
    reportNumber = "WSU-HEP-1401",
    doi = "10.1103/PhysRevD.90.074027",
    journal = "Phys. Rev. D",
    volume = "90",
    number = "7",
    pages = "074027",
    year = "2014"
}

@article{Belushkin:2006qa,
    author = "Belushkin, M. A. and Hammer, H. -W. and Meissner, U. -G.",
    title = "{Dispersion analysis of the nucleon form-factors including meson continua}",
    eprint = "hep-ph/0608337",
    archivePrefix = "arXiv",
    reportNumber = "HISKP-TH-06-25",
    doi = "10.1103/PhysRevC.75.035202",
    journal = "Phys. Rev. C",
    volume = "75",
    pages = "035202",
    year = "2007"
}

@article{Bezginov:2019mdi,
    author = "Bezginov, N. and Valdez, T. and Horbatsch, M. and Marsman, A. and Vutha, A. C. and Hessels, E. A.",
    title = "{A measurement of the atomic hydrogen Lamb shift and the proton charge radius}",
    doi = "10.1126/science.aau7807",
    journal = "Science",
    volume = "365",
    number = "6457",
    pages = "1007--1012",
    year = "2019"
}

@article{Xiong:2019umf,
    author = "Xiong, W. and others",
    title = "{A small proton charge radius from an electron{\textendash}proton scattering experiment}",
    doi = "10.1038/s41586-019-1721-2",
    journal = "Nature",
    volume = "575",
    number = "7781",
    pages = "147--150",
    year = "2019"
}

@article{Sick:2001rh,
    author = "Sick, I.",
    title = "{Elastic electron scattering from light nuclei}",
    eprint = "nucl-ex/0208009",
    archivePrefix = "arXiv",
    doi = "10.1016/S0146-6410(01)00156-9",
    journal = "Prog. Part. Nucl. Phys.",
    volume = "47",
    pages = "245--318",
    year = "2001"
}

@article{CREMA:2025zpo,
    author = "Schuhmann, Karsten and others",
    collaboration = "CREMA",
    title = "{The helion charge radius from laser spectroscopy of muonic helium-3 ions}",
    doi = "10.1126/science.adj2610",
    journal = "Science",
    volume = "388",
    number = "6749",
    pages = "adj2610",
    year = "2025"
}

@article{Braaten:2004rn,
    author = "Braaten, Eric and Hammer, H. -W.",
    title = "{Universality in few-body systems with large scattering length}",
    eprint = "cond-mat/0410417",
    archivePrefix = "arXiv",
    reportNumber = "INT-PUB-04-27",
    doi = "10.1016/j.physrep.2006.03.001",
    journal = "Phys. Rept.",
    volume = "428",
    pages = "259--390",
    year = "2006"
}

@article{Kaplan:1996rk,
    author = "Kaplan, David B. and Manohar, Aneesh V.",
    title = "{The Nucleon-nucleon potential in the 1/N(c) expansion}",
    eprint = "nucl-th/9612021",
    archivePrefix = "arXiv",
    reportNumber = "DOE-ER-40561-302, INT-96-00-156, UW-PT-96-32, UCSD-PTH-96-10",
    doi = "10.1103/PhysRevC.56.76",
    journal = "Phys. Rev. C",
    volume = "56",
    pages = "76--83",
    year = "1997"
}

@article{Kaplan:1995yg,
    author = "Kaplan, David B. and Savage, Martin J.",
    title = "{The Spin flavor dependence of nuclear forces from large n QCD}",
    eprint = "hep-ph/9509371",
    archivePrefix = "arXiv",
    reportNumber = "DOE-ER-40561-230, INT-95-00-104, UW-PT-95-14, DOE-ER-40682-102, DOE/ER/40561-230-INT95-00-104, CMU-HEP95-13",
    doi = "10.1016/0370-2693(95)01277-X",
    journal = "Phys. Lett. B",
    volume = "365",
    pages = "244--251",
    year = "1996"
}

@article{Lee:2020esp,
    author = "Lee, Dean and others",
    title = "{Hidden Spin-Isospin Exchange Symmetry}",
    eprint = "2010.09420",
    archivePrefix = "arXiv",
    primaryClass = "nucl-th",
    doi = "10.1103/PhysRevLett.127.062501",
    journal = "Phys. Rev. Lett.",
    volume = "127",
    number = "6",
    pages = "062501",
    year = "2021"
}

@article{Lin:2024bor,
    author = "Lin, Xincheng and Vanasse, Jared",
    title = "{Two-body triton photodisintegration and Wigner-SU(4) symmetry}",
    eprint = "2408.14602",
    archivePrefix = "arXiv",
    primaryClass = "nucl-th",
    doi = "10.1103/qgcx-r63r",
    journal = "Phys. Rev. C",
    volume = "112",
    number = "2",
    pages = "024001",
    year = "2025"
}

@article{Faddeev:1960su,
    author = "Faddeev, L. D.",
    title = "{Scattering Theory for a Three-Particle System}",
    journal = "Sov. Phys. JETP",
    volume = "12",
    pages = "1014--1019",
    year = "1961"
}

@article{Marcucci:2015yla,
    author = "Marcucci, L. E. and Mangano, G. and Kievsky, A. and Viviani, M.",
    title = "{Implication of the proton-deuteron radiative capture for Big Bang Nucleosynthesis}",
    eprint = "1510.07877",
    archivePrefix = "arXiv",
    primaryClass = "nucl-th",
    doi = "10.1103/PhysRevLett.116.102501",
    journal = "Phys. Rev. Lett.",
    volume = "116",
    number = "10",
    pages = "102501",
    year = "2016",
    note = "[Erratum: Phys.Rev.Lett. 117, 049901 (2016)]"
}

@article{Stockel:2024hde,
    author = {St{\"o}ckel, K. and others},
    title = "{Novel approach to infer the H2(p,{\ensuremath{\gamma}})He3 angular distribution: Experimental results and comparison with theoretical calculations}",
    doi = "10.1103/PhysRevC.110.L032801",
    journal = "Phys. Rev. C",
    volume = "110",
    number = "3",
    pages = "L032801",
    year = "2024"
}

@article{Cavanna:2018mdc,
    author = "Cavanna, Francesca and Prati, Paolo",
    title = "{Direct measurement of nuclear cross-section of astrophysical interest: Results and perspectives}",
    doi = "10.1142/S0217751X18430108",
    journal = "Int. J. Mod. Phys. A",
    volume = "33",
    number = "09",
    pages = "1843010",
    year = "2018"
}

@article{Ando:2007fh,
    author = "Ando, Shung-ichi and Shin, Jae Won and Hyun, Chang Ho and Hong, Seung Woo",
    title = "{Low energy proton-proton scattering in effective field theory}",
    eprint = "0704.2312",
    archivePrefix = "arXiv",
    primaryClass = "nucl-th",
    doi = "10.1103/PhysRevC.76.064001",
    journal = "Phys. Rev. C",
    volume = "76",
    pages = "064001",
    year = "2007"
}

@article{Kong:1999sf,
    author = "Kong, Xinwei and Ravndal, Finn",
    title = "{Coulomb effects in low-energy proton proton scattering}",
    eprint = "hep-ph/9903523",
    archivePrefix = "arXiv",
    doi = "10.1016/S0375-9474(99)00406-6",
    journal = "Nucl. Phys. A",
    volume = "665",
    pages = "137--163",
    year = "2000"
}

@article{Kong:1998sx,
    author = "Kong, Xinwei and Ravndal, Finn",
    title = "{Proton proton scattering lengths from effective field theory}",
    eprint = "nucl-th/9811076",
    archivePrefix = "arXiv",
    doi = "10.1016/S0370-2693(99)00144-6",
    journal = "Phys. Lett. B",
    volume = "450",
    pages = "320--324",
    year = "1999",
    note = "[Erratum: Phys.Lett.B 458, 565--565 (1999)]"
}

@unpublished{HaAndJared2026,
  author = {Nguyen, Ha  and Vanasse, Jared},
  title = {Coulomb Corrections to Three Nucleon Moments},
  year = {2026},
  note = {Manuscript in preparation}
}

@article{Bedaque:1999vb,
    author = "Bedaque, Paulo F. and Griesshammer, Harald W.",
    title = "{Quartet S wave neutron deuteron scattering in effective field theory}",
    eprint = "nucl-th/9907077",
    archivePrefix = "arXiv",
    reportNumber = "DOE-ER-40561-33, NT-UW-99-3",
    doi = "10.1016/S0375-9474(99)00691-0",
    journal = "Nucl. Phys. A",
    volume = "671",
    pages = "357--379",
    year = "2000"
}

@ARTICLE{2019EPJA...55..137T,
       author = {{Ti{\v{s}}ma}, Isabela and {Lipoglav{\v{s}}ek}, Matej and {Mihovilovi{\v{c}}}, Miha and {Markelj}, Sabina and {Vencelj}, Matja{\v{z}} and {Vesi{\'c}}, Jelena},
        title = "{Experimental cross section and angular distribution of the $^{2}$H(p,{\ensuremath{\gamma}})$^{3}$He reaction at Big-Bang nucleosynthesis energies}",
      journal = {European Physical Journal A},
         year = 2019,
        month = aug,
       volume = {55},
       number = {8},
          eid = {137},
        pages = {137},
          doi = {10.1140/epja/i2019-12816-1},
       adsurl = {https://ui.adsabs.harvard.edu/abs/2019EPJA...55..137T}
}

@article{Platter:2005sj,
    author = "Platter, L. and Hammer, H. -W.",
    title = "{Universality in the triton charge form-factor}",
    eprint = "nucl-th/0509045",
    archivePrefix = "arXiv",
    doi = "10.1016/j.nuclphysa.2005.11.023",
    journal = "Nucl. Phys. A",
    volume = "766",
    pages = "132--141",
    year = "2006"
}

@article{Kirscher:2015zoa,
    author = "Kirscher, J. and Gazit, D.",
    title = "{The Coulomb interaction in Helium-3: Interplay of strong short-range and weak long-range potentials}",
    eprint = "1510.00118",
    archivePrefix = "arXiv",
    primaryClass = "nucl-th",
    doi = "10.1016/j.physletb.2016.02.011",
    journal = "Phys. Lett. B",
    volume = "755",
    pages = "253--260",
    year = "2016"
}

@misc{Andis:2025fsg,
    author = {Andis, Andrew J. and Lyu, Songlin and Long, Bingwei and K{\"o}nig, Sebastian},
    title = "{Perturbative EFT calculation of the deuteron longitudinal response function}",
    eprint = "2512.12823",
    archivePrefix = "arXiv",
    primaryClass = "nucl-th",
    month = "12",
    year = "2025"
}
\end{document}